Preprinted manuscript

# Agricultural quality matrix-based multiomics structural analysis of carrots in soils fertilized with thermophile-fermented compost

Hirokuni Miyamoto*[1,2,3,4], Katsumi Shigeta[5], Wataru Suda[2], Yasunori Ichihashi[6], Naoto Nihei[7], Makiko Matsuura[1,3], Arisa Tsuboi[4], Naoki Tominaga[5], Masahiko Aono[5], Muneo Sato[8], Shunya Taguchi[9], Teruno Nakaguma[1,3,4], Naoko Tsuji[3], Chitose Ishii[2,3], Teruo Matsushita[3,4], Chie Shindo[2], Toshiaki Ito[10], Tamotsu Kato[2], Atsushi Kurotani[8,11], Hideaki Shima[8], Shigeharu Moriya[12], Satoshi Wada[12], Sankichi Horiuchi[13], Takashi Satoh[14], Kenichi Mori[1,3,4], Takumi Nishiuchi[15], Hisashi Miyamoto[3,16], Hiroaki Kodama[1], Masahira Hattori[2,17,18], Hiroshi Ohno[2], Jun Kikuchi*[8], Masami Yokota Hirai*[8]

*Affiliations:*
1. Graduate School of Horticulture, Chiba University, Matsudo, Chiba 271-8501, Japan
2. RIKEN Center for Integrative Medical Science, Yokohama, Kanagawa 230-0045, Japan
3. Sermas Co., Ltd., Ichikawa, Chiba 272-0033, Japan
4. Japan Eco-science (Nikkan Kagaku) Co. Ltd., Chiba, Chiba 260-0034, Japan
5. Takii Seed Co.Ltd., Konan, Shiga 520-3231, Japan
6. RIKEN BioResource Research Center, Tsukuba, Ibaraki 305-0074, Japan
7. Faculty of Food and Agricultural Sciences, Fukushima University, Fukushima, Fukushima 960-1296, Japan
8. RIKEN Center for Sustainable Resource Science, Yokohama, Kanagawa 230-0045, Japan
9. Center for Frontier Medical Engineering, Chiba University, Chiba, Chiba 263-8522, Japan
10. Keiyo Gas Energy Solution Co. Ltd., Ichikawa, Chiba 272-0033, Japan
11.Research Center for Agricultural Information Technology, National Agriculture and Food Research Organization, Tsukuba, Ibaraki, 305-0856, Japan
12. RIKEN, Center for Advanced Photonics, Wako, Saitama, 351-0198, Japan
13. Division of Gastroenterology and Hepatology, The Jikei University School of Medicine, Kashiwa Hospital, Kashiwa, Chiba 277-8567, Japan
14. Division of Hematology, Kitasato University School of Allied Health Sciences, Sagamihara, Kanagawa 252-0329, Japan
15. Division of Integrated Omics research, Bioscience Core Facility, Research Center for Experimental Modeling of Human Disease, Kanazawa University, Kanazawa, Ishikawa, 920-8640, Japan
16. Miroku Co.Ltd., Kitsuki, Oita 873-0021, Japan
17. School of Advanced Science and Engineering, Waseda University, Tokyo169-8555, Japan
18. School of Agricultural and Life Sciences, The University of Tokyo, Bunkyo, Tokyo 113-8657, Japan

* Hirokuni Miyamoto Ph.D. *Graduate School of Horticulture, Chiba University, RIKEN Center for Integrative Medical Science; Sermas Co., Ltd.; Japan Eco-science Co. Ltd.*
**Email:**  hirokuni.miyamoto@riken.jp
* Jun Kikuchi, Ph.D. *RIKEN Center for Sustainable Resource Science*
**Email:**  jun.kikuchi@riken.jp
* Masami Yokata Hirai, Ph.D. *RIKEN Center for Sustainable Resource Science*
**Email:** masami.hirai@riken.jp

**Abstract**
Compost is used worldwide as a soil conditioner for crops, but its functions have still been explored. Here, the omics profiles of carrots were investigated, as a root vegetable plant model, in a field amended with compost fermented with thermophilic Bacillaceae for growth and quality indices. Exposure to compost significantly increased the productivity, antioxidant activity, red color, and taste of the carrot root and altered the soil bacterial composition with the levels of characteristic metabolites of the leaf, root, and soil. Based on the data, structural equation modeling (SEM) estimated that L-2-aminoadipate, phenylalanine, flavonoids and / or carotenoids in plants were optimally linked by exposure to compost. The SEM of the soil estimated that the genus *Paenibacillus*, L-2-aminoadipate and nicotinamide, and S-methyl L-cysteine were optimally involved during exposure. These estimates did not show a contradiction between the whole genomic analysis of compost-derived *Paenibacillus* isolates and the bioactivity data, inferring the presence of a complex cascade of plant growth-promoting effects and modulation of the nitrogen cycle by compost itself. These observations have provided information on the qualitative indicators of compost in complex soil-plant interactions and offer a new perspective for chemically independent sustainable agriculture through the efficient use of natural nitrogen.

*Abbreviations*: SEM, structural equation modeling; SDGs, sustainable development goals

## Introduction

Global food shortages are an urgent issue, and malnutrition remains a major cause of death in some areas. In addition, because deficiencies in trace components cause many diseases, it is necessary to develop innovative agricultural technologies to increase crop production and nutritional value [1-3].

The planetary boundary framework [4] has emphasized the deterioration of biodiversity and available nitrogen and phosphorus. Traditionally, chemical fertilizers have been considered essential and distributed in agriculture, but this approach has increased the burden on ecosystems [5,6]. Furthermore, recent studies have shown that amino acids are more critical to plants than inorganic nitrogen [5]. In this context, the importance of organic farming should be reconsidered [7,8]. Recycling agriculture is essential to meet the goal of sustainable development by efficiently using nitrogen and phosphorus [8,9]. However, there is a history of contention about how the quality of organic compost affects crops in many respects [10]. Furthermore, the roles of compost in different stages of an ecosystem appear to depend on the environmental conditions [8,11,12]. In general, the quality of composts seems to be unstable because several different types of raw materials are fermented under uncertain fermentation conditions within composts [13-15], e.g., the contents of fermentation bacteria and moisture and other conditions of the fermentation process may vary. Therefore, although there is an assumption of the effectiveness of compost in resource recycling, it contains uncertainty due to the unstable characteristics of compost fermentation.

Studies on thermostable and thermophilic *Bacillus* species, which may be involved in stable composting and plant health contributors as plant growth-promoting bacteria (PGPB) candidates, have been reported[16-18]. Miyamoto, Kodama, and their joint research group have reported PGPB with antifungal activity [19] in a compost model: it was produced at high temperature (over 70 °C as spontaneous fermentation) with marine animal resources (MAR compost) in a fed-batch system of bioreactors containing a thermophilic Bacillaceae as stable fermentation bacterial community[19]. Interestingly, the compost had a strain that played the role of PGPB with antifungal activity [19] and reduced the accumulation of plant nitrate by denitrifying activity in soils [20]. Furthermore, oral administration of compost or its extract and the compost-derived thermophile Bacillaceae can improve the fecundity and quality of rodent animal models, fisheries and livestock animals [21-27]. Under *in vitro* anaerobic conditions, Tashiro and Sakai *et al.* demonstrated that compost and its derived thermophilic *Bacillus* efficiently produced optically pure L-lactate (100% optical purity) from nonsterilized kitchen refuse [28-31] or starch as a material [30]. These observations, including the findings of our research groups, suggest that Bacillaceae in compost may stably change the structure and function of symbiotic bacteria and their habitats.

Here, we aim to assess compost-soil-plant interactions using carrots. This crop model allows easy analysis of portions above and below the ground after exposure to this thermophilic Bacillaceae-fermented compost and to explore the active bacterial groups. Multiple omics analysis has been used as an innovative method for various research objects [5,32]. Using integrated analyses of multiple omics datasets, including plant and soil metabolome and microbiome (Fig. S1), we classified components linked to carrot productivity, color matrix, and taste indices. The covariance structural analysis with the classified components statistically estimated the multiple regression models between plant growth indices, plant and soil metabolites, and the soil microbiota. Based on the regression model, it was inferred that *Paenibacillus* strains as nitrogen-fixing bacteria are involved in the effect of compost. After these estimates, the isolation, genome analysis, and evaluation of the biological activity of compost-derived *Paenibacillus* strains were advanced. As a result, the possibility that the calculation hypothesis was properly taken into account was anticipated. Using these computational procedures to advance biochemical experiments has effectively assessed the complex relationship between soil, plants, and compost. The information from this study could provide an essential perspective for the construction of sustainable agricultural technologies in the future.

## Materials and Methods

Carrot seeds (Takii Seeds Phytorich Series, Kyo Kurenai, Takii Seeds Co., Ltd.) were sown in two rows located 5 cm apart on the cultivation area, and the rows were covered with nonwoven fabric after sowing. The plants were harvested twice, in November and February, and their stem and leaf weight, root weight, root diameter, root length, color, taste, the metabolites, and anti-oxidant activities were measured. Soil was collected after harvesting, the metabolites and bacterial populations were investigated. On the basis of the omics data, correlation analyses, association analyses [33,34] were performed. Furthermore, covariance structure analysis/structural equation modeling (SEM), causal mediation analysis (CMA), and BayesLiNGAM were performed as previously described [34]. Based on the predicted model, functional bacterial candidates from compost were isolated, genomic analysis was performed, and biological activity was measured. The F test of the data was used to select parametric and non-parametric analyses. The unpaired t-test and Wilcoxon test were selected as appropriate methods depending on the data sets. Significance was declared when $P < 0.05$, and a trend was assumed at $0.05 \leq P < 0.20$. These calculated data were prepared using R software and Prism software (version 9.1.2) and were visualized using an estimation plot or entering replicates in a data graph. Data are presented as the means ± SEs. Detailed testing methods not listed here are described in the Supplementary methods.

## Results
### Harvest survey
The experiment in which the soil was fertilized with thermophile-fermented compost was planned as shown in Fig. 1a. The amendment with compost tended to increase the weight of the stems, leaves, and roots of carrots. This tendency showed marked increases dependent on the duration of the exposure to the compost (Figs. 1b and 1c). A significant difference was observed in February (Fig. 1c). Interestingly, the carrots fertilized with compost tended to turn red. Therefore, an analysis was performed using the RGB color matrix as an index according to the procedure shown in Fig. 1d. The study of the color matrix of randomly sampled carrots showed that the rate of red coloration was significantly increased by compost amendment ($p=0.012$) (Fig. 1e). Because carotenoids generally cause the red color of carrots, the carotenoid concentrations were examined. The results confirmed that the concentrations of α-carotene ($p=0.0103$ vs. control), beta-carotene ($p=0.1188$ vs. control), and lycopene ($p=0.0146$ vs. control) tended to increase (Fig. S2). Furthermore, a taste survey indicated apparent differences in flavor richness and immaturity. These differences were confirmed in the carrots harvested in February (Fig. 1f) and those harvested in November, which showed a slight effect on growth (Fig. 1b). These observations confirmed that the compost amendment improved the indices of carrot production and quality in this study.

### Metabolome analyses of carrot leaves and roots
The application of thermophile-fermented compost increased the content of 2,2-diphenyl-1-picrylhydrazyl (DPPH) content by 40% ($p=0.2612$) in the roots. At the same time, it decreased the DPPH content by 40% in the leaves ($p=0.0014$) (Fig. 2a). On the basis of these evaluations, metabolome analyses of carrot leaves and roots were performed. Additionally, correlation analysis was performed on the basis of the whole analyzed data. As a result, the correlations of amino acids, flavonoids, and phenylpropanoids were different between the control and compost groups (Fig. S3).

In detail, the levels of metabolic compound candidates changed as follows: the levels of 4-amino-butyric acid (GABA), as amino acid-related metabolites, increased significantly in the leaf after the compost exposure ($p=0.0072$), but methionine sulfoxide and methionine levels were decreased considerably ($p=0.0007$). The levels of the metabolites annotated as kaempferol-Gal-Rha, cyanidin 3-O-Rut, apigenin 7-O-neohespseridoside, and quercetin-Glc as flavonoids in the leaf were significantly decreased by exposure ($p=0.0023$; $p=0.0031$; $p=0.0035$; and $p=0.0147$, respectively) (Fig. 2b). For other categories, indole 3-carboxyaldehyde increased considerably by exposure ($p=0.009$), but L-2-aminoadipate, malate, 1,3-dimethylurate, methylmalonate, and phenyllactate decreased ($p=0.0079$; $p=0.0106$; $p=0.0230$; $p=0.0410$; and $p=0.0461$, respectively). In roots (Fig. 2c), tryptophan, phenylalanine, tyrosine, and L-2-aminoadipate levels, as amino acids and nitrogen metabolites, were significantly reduced by the exposure ($p=0.0177$; $p=0.0952$; $p=0.0449$; and $p=0.0355$, respectively), and arginine levels were significantly increased ($p=0.0315$). The correlation between these individual metabolites was also clearly different (Fig. S4). Thus, such an increase or decrease in common metabolites was not necessarily confirmed. However, at least the compost amendments significantly changed leaf and root metabolites, especially leaf flavonoids involved in antioxidant activity. It has become possible to classify characteristic metabolites in carrot leaves and roots depending on the cultivation conditions.

### Soil omics analysis
To evaluate the relationships among leaves, roots, and soils, soil omics analyses were also performed. Due to the difference in the target and analysis conditions, the metabolome analysis was performed using a device different from the one used for the leaf and root. Of the amino acids and related compounds, *S*-methyl-L-cysteine had an increased tendency in the composting group, and the levels of nicotinamide tended to decrease (Fig. 2d).

Subsequent analysis of the soil bacterial population suggested that the bacterial diversity differed between conditions with and without the application of compost (Fig. S5a and Fig.3a). There was little change in α-diversity (Fig. S5a). However, the beta diversity tended to change, although not significantly (Fig. 3a). Some changes in the bacterial populations were observed that were not always significant differences (Figs. 3b and S6a). Evaluation at the phylum level showed that the phylum Proteobacteria as a dominant bacterial group decreased significantly after the compost amendment. The levels of the phyla Gemmatimonadetes, Verrucomicrobia, Firmicutes, Elusimicrobia, and Thaumarchaeota tended to be high (Figs. 3b and S6a). On the contrary, the phyla Planctomycetes, Acidobacteria, Spirochaetes, and Deinococcus-Thermus tended to decrease. At the genus level (Figs. 3c and S6b), the abundance of the genus *Marmoricola* increased significantly. Furthermore, the genera *Paenibacillus*, *Geobacillus*, *Panacagrimonas*, *Nocardiodes*, and *Phycicoccus* tended to increase.

Therefore, these observations suggested that the amendment of the compost caused changes in the bacterial populations of the soil and altered the metabolism of the soil.

### Association analyses and covariance structure analyses
An association analysis was performed based on production indices, color matrix, and omics data, and factors strongly related to compost administration were selected (Fig. S7). The characteristics of leaf, root, and soil bacteria were also visualized, and the differences in the relationships differed from the statistically significant differences in the individual factors (Fig. S8). They also contained relatively low concentrations of metabolites and minor bacteria. Based on the classified data, structural equation analysis was carried out to understand the basic structure of the skeleton in this complex system.

The structural equation calculates the combination showing the optimum value. Among the calculated models for DPPH as an indicator of antioxidant activity, the multiple regression model with amino acids, flavonoids, and carotenoids (Figs. 4a and S9a) had the highest optimal fit index values (Table S1). Therefore, the structural equation for flavonoids was calculated with phenylalanine, a flavonoid precursor, focusing on the factors that actually changed in the above observations. As a result, apigenin 7-O-neohesperidoside, kaempferol-Gal-Rha, and quercetin_Glc, as leaf flavonoids, root L-2-aminoadipate, and phenylalanine formulated the optimal structural equation (Figs. 4b and S9b) rather than the other inferior models (Table S2). These estimations suggested that compost amendments are structurally involved as a group in the concentration of flavonoids, amino acids, and carotenoids in carrots, as well as in the antioxidant activity of carrots.

Next, the relationships between the changes observed in the compost-treated soil were constructed by modeling structural equations. The soil bacterial candidates and soil metabolites that showed representative changes by compost amendment were selected in the above experiment. The relationship of the soil factors associated with compost amendments was visualized as optimal SEM (Figs. 4c and S9c): an optimal model that included the genus *Paenibacillus*, DL-2-aminoadipate, nicotinamide, and S-methyl L-cysteine showed good fit indices compared to those of the other models (Table S3). Furthermore, causal mediation analysis revealed that a single mediation (indirect) effect (ACME) was not calculable or without significant values (Tables S4, S5, and S6). These calculated results estimated a strong relationship as each group. Therefore, the causal interactive relationships within each optimal model of the leaf, root, and soil were investigated based on a calculation by BayesLiNGAM (Figs. S10). The highest causal interactive relationships (top six) were mainly shown from compost but not always. This may mean that the compost supports it while the physiological responses of the carrot and/or the soil themselves are prioritized. These results pointed to the importance of the optimal model as a whole group, as well as the optimal model group estimated for the leaf and root.

Therefore, nitrogen compounds were involved in these structural equations for leaves, roots, and soil. Based on these observations, the genus *Paenibacillus* derived from compost was explored, followed by a search for its function within the compost itself.

### Genome profiles associated with plant growth promotion

As described above, the SEM results suggested that the genus *Paenibacillus* may be a necessary component of compost for the promotion of plant growth. Therefore, we isolated compost-derived colonies that tested positive for a nitrogen fixation gene and were able to select two strains that had a nitrogen fixation reaction from compost-cultured colonies. The compost-derived strains *Paenibacillus macerans* HMSSN-036 and *Paenibacillus* sp. HMSSN-139 were identified, and their genomes were analyzed: the former, a strain closely related to *Paenibacillus macerans* (identity 99.0%); the latter, a candidate as a new species belonging to *Paenibacillus*. In Figs. 5a-d, electron micrographs of these strains showed the spore- and vegetative-forms. The molecular phylogenetic tree of these isolates is shown in Fig. 5e. Phylogenetic analysis with mash distance (Fig. S11) showed that the isolated *Paenibacillus* strains were classified into weak sequences for *P. curdianolyticus* and *P. kobensis* detected by meta-sequence analysis of the 16S bacterial rRNA gene sequence: Their population was not detected in the control and 0.022± 0.011% in the test group (p=0.116). These observations indicated that the isolated *Paenibacillus* strains did not necessarily coincide with the sequence of *Paenibacillus* species increasing in the compost-amended soil. Still, it is interesting that the same *Paenibacillus* increased in the soil. Furthermore, genome analyses showed the presence of the genes for nitrogen fixation, auxin production, phosphate solubilization, and siderophore reactions such as PGPR (Tables 1 and S7-S10). Based on genomic information, biological assays were performing for auxin production, siderophore reaction, and phosphate solubilization and all of these assays were demonstrated to be positive (Fig. S12). Therefore, the genes detected in isolated *Paenibacillus* strains could be functional for nitrogen fixation, auxin production, siderophore reaction, and phosphate solubilization in the bacterium.

These results suggest that compost-derived *Paenibacillus* strains may exhibit PGPR effects.

### Evaluation of the effects of compost itself on the promotion of plant growth

The nitrogen cycle function of compost with nitrogen-fixing *Paenibacillus* was then evaluated with *in vitro* soil models. Two functional evaluations of nitrogen fixation and reduction of nitrous oxide generation were conducted.

First, as shown in Fig. S12a, the nitrogen fixation capacity was evaluated in soil with the stable nitrogen isotope $^{15}N$ added. The *Arabidopsis thaliana* model plant was cultured to assess the effects on nitrogen fixation, and a growth promotion effect, although not significant, was observed (Fig. S12b). Under such conditions, nitrogen fixation rates of approximately 20% were possible in the plant and soil. Based on these observations, it is possible that the soil modulates nitrogen fixation in the field after exposure to compost. The results of the elemental analysis of the soil immediately after the cultivation of the carrot-grown soil in this field test changed significantly, although it was tested after it was well mixed with the soil in the control and test areas prior. Total nitrogen tended to increase slightly, and the EC value related to the nitrate concentration was higher in the test area (Table S11).

Next, a system was created to measure the amount of $N_2O$ generated from the soil (Fig. S12c). The results of the NGS analysis of the prepared soil indicated that it mainly contained *Gibberella* sp. (66.1%) and *Fusarium* sp. (28.7%) as fungi. The potato dextrose power, which can generally be used as a medium for fungi, was added to the soil for the test. The test results confirmed that the conditions with added compost showed markedly reduced $N_2O$ generation compared to that of the control group (soil only with fungi and PDA) (Fig. S12d). Since $N_2O$ content was approximately the same as those under the condition with no nutritional source of fungi as a reference test, it was suggested that the inhibitory effect was even greater. These results are based on the test laboratory level. In contrast, measuring nitrous oxide at the field level and exploring the mechanism of action of complex microorganisms is costly and time-consuming. At any point, the phenomena identified in this study are of social significance. On the contrary, the use of such a laboratory system suggests that it is possible to assess the potential for $N_2O$ generation in soil collected from the field in the laboratory without measuring the volume of $N_2O$ at the field level.

Furthermore, it should be noted that the soil tested immediately after harvesting the carrot fields confirmed an increase in PAC and a decrease in iron concentration (Table S11). The field test results were consistent with the compost-derived *Paenibacillus* strains being positive for phosphate solubilization and siderophore reactions. Furthermore, the growth of carrots in this field experiment was not contradictory, even though the isolated *Paenibacillus* strains from the compost were positive for producing the plant hormone auxin.

Thus, the function of the compost was verified in the field test and had similarities with the results of the *in vitro* test as complementary data.

**Discussion**

Here, we succeeded in deriving structural equations related to nitrogen metabolism in leaves, roots, and soil as the effect of thermophile-fermented compost. In the structural equation, it was assumed that *Paenibacillus*, a candidate for nitrogen fixation, might be involved, and we succeeded in isolating *Paenibacillus* strains derived from the compost. Furthermore, based on genomic analysis and evaluation of biological activity, the possibility of nitrogen circulation in plants and soil by the compost used here was confirmed.

In plants, the structural equations of amino acids and flavonoids were inferred. The optimal compost-linked SEM linked with apigenin 7-O-neohesperidoside, kaempferol-Gal-Rha, and quercetin Glc as leaf flavonoids was consistent with that phenylalanine being the precursor of flavonoids [35,36]. The model provided a new insight on using L-2-aminoadipate as a nitrogen compound. The 2-aminoadipate candidates commonly detected in leaves, roots, and soil were linked to the estimated SEM candidates (Tables S2 and S3). L-2-aminoadipate was significantly altered in both leaves and roots with the addition of compost, but DL-2-aminoadipate was not always changed in the soil. Furthermore, the difference in the L and DL types may depend on different equipment, and the cause was unclear. In any case, the estimated SEM candidates with 2-aminoadipate had relatively suitable fit index values. Since 2-aminoadipate is associated with the activity of lysine–oxoglutarate reductase [37], the relationship with lysine was reassessed. 2-Aminoadipate was not only reduced in leaves and roots (Fig. S13a), but the proportion in leaves and roots increased conversely in the roots (Fig. S13b). Lysine tended to increase in roots (Fig. S13c), and the proportions in leaves and roots also tended to increase in roots (Fig. S13d). Root lysine was not adopted as a factor in the structural equation in Fig. 4b because it was not significant ($p < 0.05$, Fig. 2c) and did not have a high lift value in the association analysis (Fig. S7). However, the structural equation, including lysine, was calculated on the basis of the information on physiological function. As in results (Table S12), the model that included some regression model formulas showed the optimum value, along with the models with a metabolite annotated as metformin and arginine, rather than the model that contained lysine and aminoadipate. In particular, metformin and arginine also increased in the roots (Fig. 2c). Metformin is a component of therapeutic agents for the treatment of diabetes [38] and an anti-aging therapy candidate [39]. Previous reports of plant-derived metformin are notable [40,41]. Arginine is potentially therapeutic for cardiovascular disorders [42]. These functional titers need to be recognized with more care that considers their physiological efficacy. However, it may make sense that nitrogen metabolites have been detected after compost amendment with nitrogen-fixing *Paenibacillus*.

In the structural equation of leaves and roots, carotenoids appeared to be directly affected by compost, but the relationship between other metabolites was not statistically clarified. In particular, the soil sampled immediately after carrot cultivation showed reduced iron (Table S11), and compost-derived *Paenibacillus* may be involved in exhibiting a siderophore response to promote iron absorption. In addition, it was reported that the biosynthesis of carotenoids was controlled by cytochrome P450 [43], which could have heme with iron as an essential component [44].

In the structural equations between soil metabolites, soil bacteria, and compost, the decreasing trend of nicotinamide and the increasing trend of *S*-methyl L-cysteine deserve attention. Nicotinamide has been reported to be involved in plant growth [45-47], biological defense[48], and antioxidant activity[49]. *S*-methyl L-cysteine sulfoxide, a derivative of *S*-methyl L-cysteine, is a repellent against plant parasitic nematodes [50], which reduce plant growth and productivity. *S*-methyl L-cysteine sulfoxide, a precursor of soil fumigant dimethyldisulide (DMDS) (Dangi et al., 2014), is produced by plants (Friedrich et al., 2022) and bacteria

(Joller et al., 2020). Although DMDS is a soil fumigant that controls soil-borne pathogens and nematodes, it has been pointed out that it may not function depending on the microbial environment of the soil (Dangi et al., 2014). In addition, some *Paenibacillus* strains are known to have a defensive effect on nematodes [51,52]. In fact, the thermophile-fermented compost used in this experiment was less likely to cause nematode damage in another field test and made it difficult to form root knots as a detriment (Fig. S14). These results appeared to be different from those involving nematodes already reported [53], although under the other experimental conditions. Nevertheless, the commonalities here may confirm a novel aspect of the defense against root-knot nematode damage using the *Paenibacillus*-harboring compost in the structural equation.

As previously reported, the 16SrRNA sequence of the genus *Paenibacillus* was one of the bacterial genera once was found in compost [19]. In this study, two strains of *Paenibacillus* spp. from compost were newly isolated as nitrogen-fixing bacteria. Genome data was consistent with the structural equations and production results in this experiment, although not all. Except for the genes for nitrogen fixation, auxin production, phosphate solubilization, siderophore reaction, and related function, genes related to GABA and/or isopropylmalate production (Table S9) were notable. GABA plays a crucial role in plant drought tolerance [54] and pathogen and insect attacks [55] and has been suggested to play a vital role in nitrogen fixation in seagrass [56], even though it is not a terrestrial plant. 2-Isopropylmalate synthase (IPMS) is involved in the synthesis of 2-isopropylmalate [57]. Notably, IPMS is involved in the biosynthesis of the amino acid leucine [58] and flavor compounds in apple. Although the plant species are different, they do not contradict the results in this study. In any case, although GABA was detected in plants, the abundance of GABA in the soil was extremely low. Furthermore, the abundance of 2-isopropylmalate in the soil was also extremely low, resulting in difficult discussions based on concentration in the soil.

The identical relationship between the genus *Paenibacillus* detected in soil omics analysis (Fig. 3c) and isolated *Paenibacillus* strains derived from compost (Table 1) should be confirmed by mash distance analysis (Fig. S11). Unfortunately, it was impossible to verify that the isolated *Paenibacillus* strains from the compost were consistent with the sequences detected in the carrot soil. However, it was interesting that the abundance of bacteria belonging to the genus *Paenibacillus* found in compost also increased in compost-treated soil. Quorum sensing is commonly defined as a signaling network between bacteria that regulates their function as a bacterial group depending on the density of a given species [59], and notably, known species belonging to *Paenibacillus* [60] and *Bacillus* [61] carry quorum sensing genes. Thus, it may be suggested that this mechanism might contribute to plant growth and that closely related *Paenibacillus* species detected in the soil from the carrot experiment play a role in the whole group of genus.

This study evaluated the impact of *Paenibacillus*-harboring compost on the nitrogen cycle in the soil, which cannot be carried out in field trials. As a result, nitrogen fixation capacity and reduction in $N_2O$ were confirmed. In recent years, it has been suggested that *Paenibacillus polymyxa* may be involved in the suppression of $N_2O$ generation[62], but the results of this study confirmed the effect of the compost itself. The $N_2O$ in the soil is generated by cytochrome P450 derived from fungi [63,64], which is often a pathogen of soil origin. Notably, a bacterium producing cyclic lipopeptides that suppress growth in fungi [19] was also present in the compost used in this experiment. Thus, thermophile-fermented compost is expected to be one of the potential candidates for ecological biostimulants. Following this research, it should be necessary to explore the conditions under which compost can be used more efficiently.

Fig. 6 shows an inference model obtained from the results of this survey. It shows that the introduction of compost along with *Paenibacillus* affects the nitrogen cycle of the soil and the plant body inherent in the soil. Expressly, the model assumes that nitrogen fixation and associated suppression of nitrous oxide production occur in the soil and that the absorption of phosphate, iron, and nicotinamide, which are plant nutrient sources, is enhanced. As a result, increases or decreases in amino acids, flavonoids, and carotenoids involved in the nitrogen cycle would be expected to be affected. Furthermore, an environmentally friendly mechanism is assumed by increasing the content of *S*-methyl L-cysteine, which has antipathogenic properties.

There is a global need for sustainable organic farming. In this study, compost fermented at high temperatures was used as a model organic fertilizer. Integrated omics analyses were conducted using carrots as a crop model, allowing easy analysis of the portions above and below the ground. Based on the characteristics of omics data, SEM estimated the optimal models that the compost itself and/or the influenced rhizosphere improve the productivity and quality of the crop and environmental conditions, such as nitrogen fixation and denitrification with reduced greenhouse gases. These statistical inferences provide a novel perspective on the potential use of sustainable compost.

**Acknowledgments** Special thanks to Mr. Shigeo Enomoto and the Miyakoji farm members (Miyakojino Tamago Co., Ltd. and Miyakoji Farm Co., Ltd.) for their appropriate advice and preparation of the experimental fields. We are also grateful to Eri Ochiai for the data arrangement, Sanyo Trading Co., Ltd., for the $N_2O$ measurements, Naoko Tachibana (RIKEN IMS) for preparing the samples, and Mr. Toru Harada (Harada Agrobusiness Co., Ltd.) and Mr. Yuji Katayama (Econet Co., Ltd.) for cucumber field management and providing photos of nematode-damaged and nondamaged cucumber.

**Figure and Table legends**

**Fig. 1** (a) Cultivation schedule. (b) Evaluated growth indices of carrots. (c) Photograph of harvested carrots. (d) The procedure for color assessment. (e) Values of RGB pixel indices in the tested carrots. (f) Results of taste evaluation. The asterisks show as follows: *, p<0.05; **, p<0.01.

**Fig. 2.** The antioxidant activity in the leaves and roots of carrots and heatmaps of correlation between the categories belonging to metabolite candidates. (a) DPPH (2,2-diphenyl-1-picrylhydrazyl) activities in the leaf and root of the carrot (n=5). The "Control" and "Compost" show the data under normal conditions (control group) and compost-amended conditions (compost group), respectively. (b)(c)(d) Metabolic differences in the leaf, root, and soil. Heatmaps show the relative abundances of the metabolites in each cluster: the control (blue) and test groups (compost-amended group)(red).The representative metabolites in the leaf and the root in each group are shown (comparison of the categories belonging to metabolites with p < 0.1 as significant values). The metabolites in the soil are shown (comparison of the categories belonging to metabolites with p < 0.2 as significant values). Significan t values were shown as follows: *, p<0.05; and **, p<0.01.

**Fig. 3.** (a) UniFrac graph (unweighted and weighted) showing β-diversity in the control and test groups. The estimation plot of the bacterial population in the soil of the control and test groups. Relative abundances of the (b) phyla and (c) genera (p <0.1; >0.1% as the maximum of bacterial population).

**Fig. 4.** The relationship of the leaf and root metabolite candidates associated with compost amendment was visualized by structural equation modeling in the regression groups selected in Tables S1 and S2. Standardized β coefficients are reported. The green and red colors show positive and negative interactions, respectively. (a) shows the sempath linked with amino acids, carotenoids, flavonoids, DPPH activity, and compost. (b) shows the sempath linked with apigenin 7-O-neohesperidoside, kaempferol-Gal-Rha, and quercetin Glc as flavonoids in the leaf and L-2-aminoadipate and phenylalanine in the root and compost. (c) shows the sempah linked with soil metabolites, *Paenibacillus*, and compost.The abbreviations are as follows: Cmp, compost; L_AA, total amino acid contents in the leaf; L_Fav, total flavonoid contents in the leaf; L_DPP, DPPH activities in the leaf; R_Car, total carotenoid contents in the root; L_Apig, apigenin 7-O-neohesperidoside in the leaf; L_Kae, kaempferol-Gal-Rha in the leaf; L_Que, quercerin Glc in the leaf; R_L2A, L-2-aminoadipate in the root; R_Phe, phenylalanine in the root; Pnb, *Paenibacillus*; S_A, L-2 aminoadipate; S_N, nicotinamide; S_M, *S*-methyl L-cysteine. The fit indices are shown in the path as follows: Chiq, chi-square $\chi^2$; p value, p values (chi-square); CFI, comparative fit index; TLI, Tucker–Lewis index; RMSEA, root mean square error of approximation; SRMR, standardized root mean residual; GFI, goodness-of-fit index; AGFI, adjusted goodness-of-fit index.

**Fig. 5.** Electron micrograph and phylogenetic relationship of isolated Paenibacillus strains in this study. (a) Scanning electron microscopy image of vegetative typed-*Paenibacillus macerans* HMSSN-036 stain. (b) Scanning electron microscopy image of spore typed-*Paenibacillus macerans* HMSSN-036 stain. (c) Scanning electron microscopy image of vegetative typed-*Paenibacillus* sp. HMSSN-139 stain. (d) Scanning electron microscopy image of spore typed-*Paenibacillus* sp. HMSSN-139 strain. (e) Whole-genome phylogenetic analysis of isolated *Paenibacillus* strains was performed. A phylogenetic tree was constructed by neighbor-joining method. The isolated *Paenibacillus* strains are marked in red.

**Fig. 6.** A putative model consequential in this study.

**Table 1.** Functional genes identified based on the genomic data of *Paenibacillus* sp. isolated from the thermophile-fermented compost (see the supplementary information).

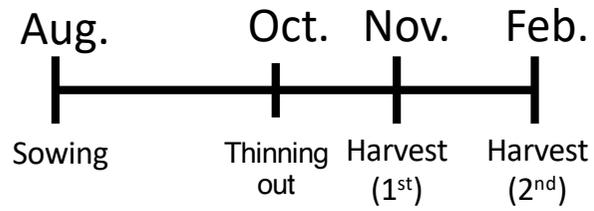
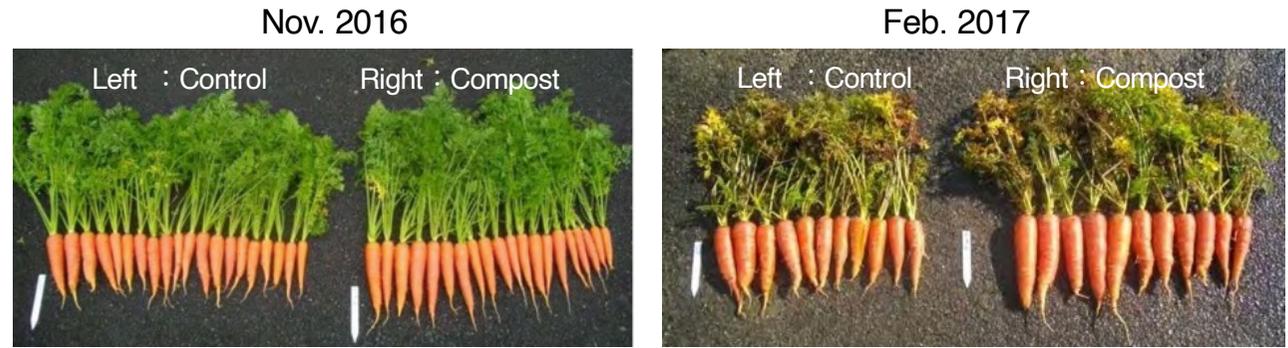

**a** Aug. Sowing — Oct. Thinning out — Nov. Harvest (1st) — Feb. Harvest (2nd)

**b** Nov. 2016 / Feb. 2017 — Left: Control, Right: Compost

**c**

| Index | 1st Nov 2016 (n=15) | | 2nd Feb 2017 (n=11) | |
|---|---|---|---|---|
| | Control | Compost | Control | Compost |
| Height (cm) | 43.8 ± 4.4 | 46.5 ± 3.5 | 40.5 ± 3.0 | 43.4 ± 6.0 |
| Root length (cm) | 16.4 ± 1.5 | 19.1 ± 1.8 * | 18.5 ± 2.0 | 21.4 ± 3.0 ** |
| Root diameter (cm) | 3.9 ± 0.5 | 4.0 ± 0.4 | 4.9 ± 0.5 | 5.4 ± 0.5 * |
| Stem & leaf weight (g) | 35.0 ± 14.0 | 34.0 ± 11.0 | 18.0 ± 7.0 | 29.0 ± 13.0 * |
| Root weight (g) | 117.0 ± 38.0 | 140.0 ± 41.0 | 201.0 ± 67.0 | 291.0 ± 97.0 * |

**f**

| Index | Nov. 2016 (n=4) | | Feb. 2017 (n=8) | |
|---|---|---|---|---|
| | Control | Compost | Control | Test |
| Sweet | 4 | 4 | 8 | 8 |
| Rich taste | 0 | 4 | 0 | 8 |
| Immaturity | 4 | 0 | 8 | 0 |
| Flavor | 0 | 4 | 0 | 8 |

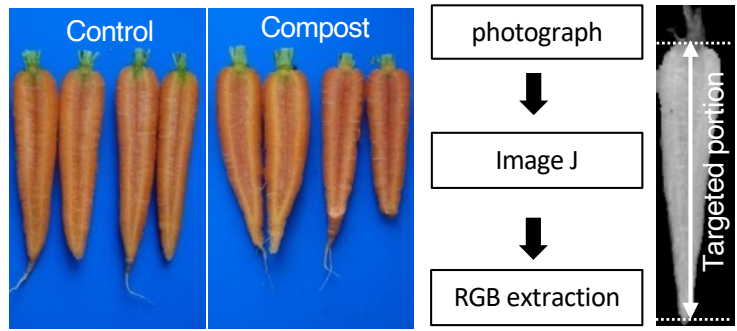
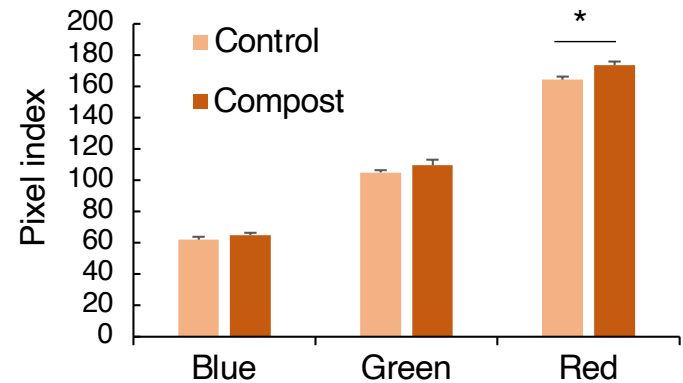

**d** photograph → Image J → RGB extraction; Targeted portion

**e** Pixel index: Blue, Green, Red (Control vs Compost) *

Fig.1

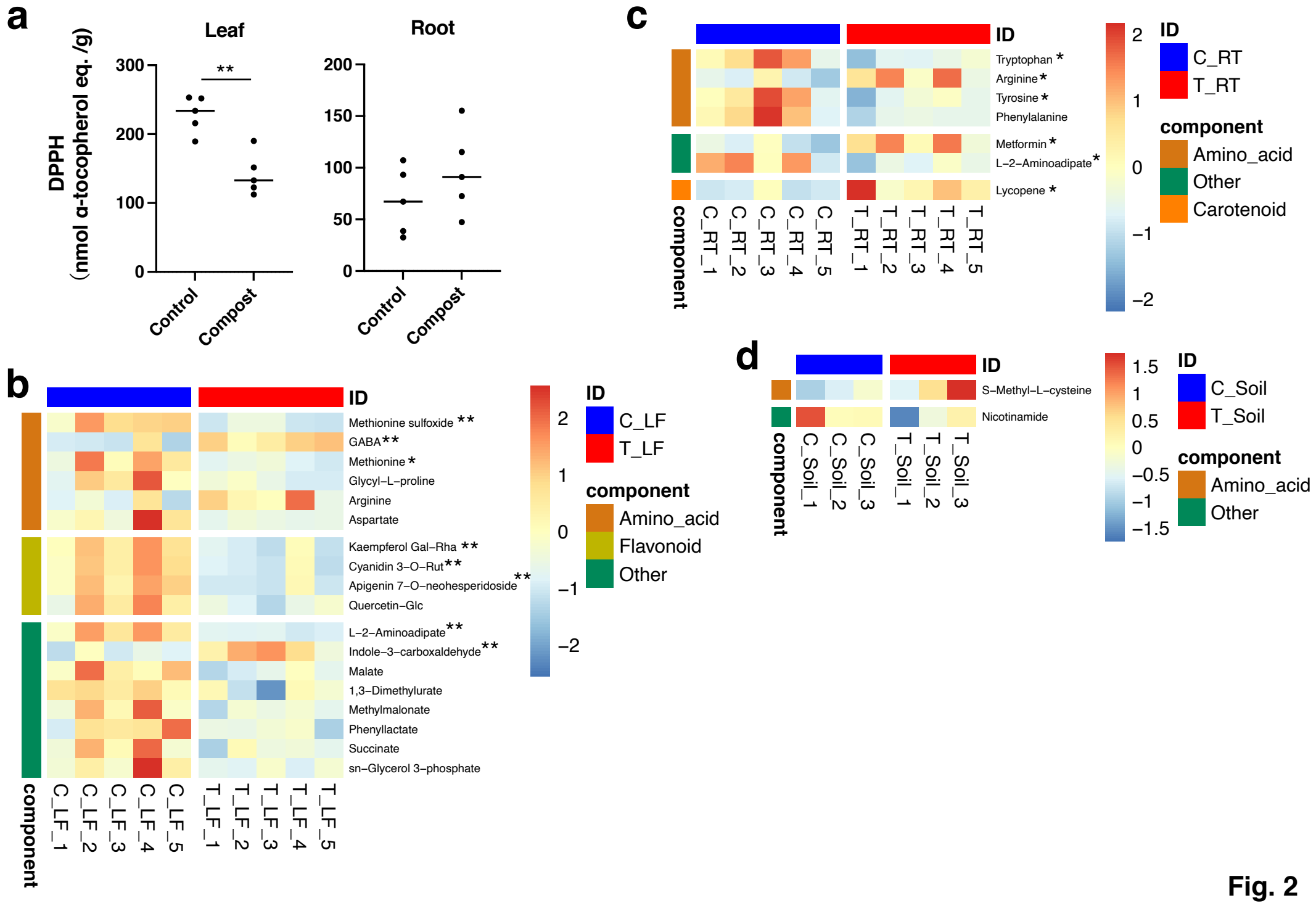

Fig. 2

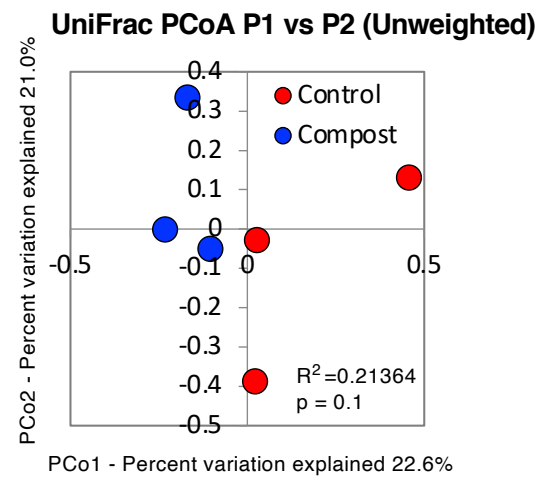
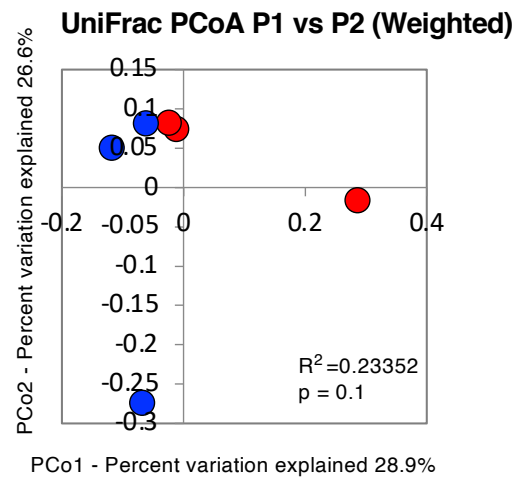
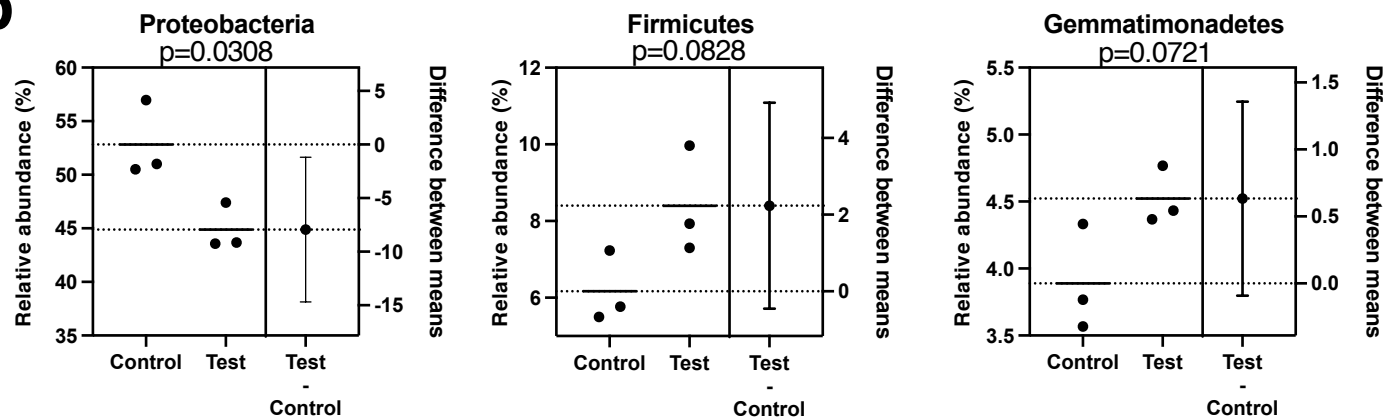
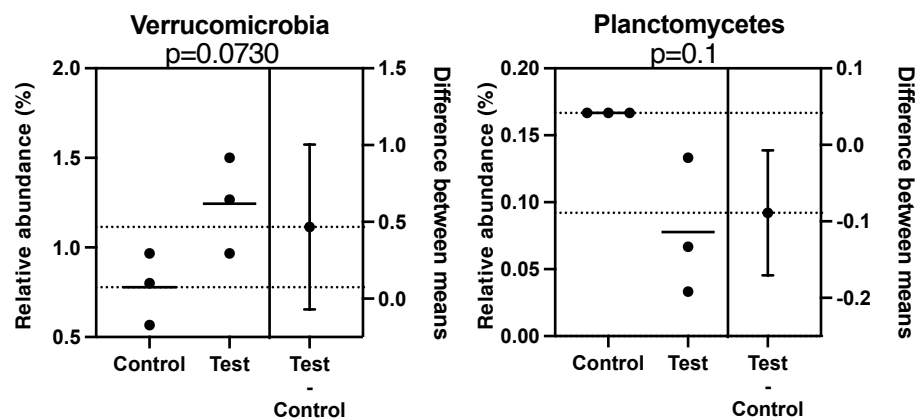
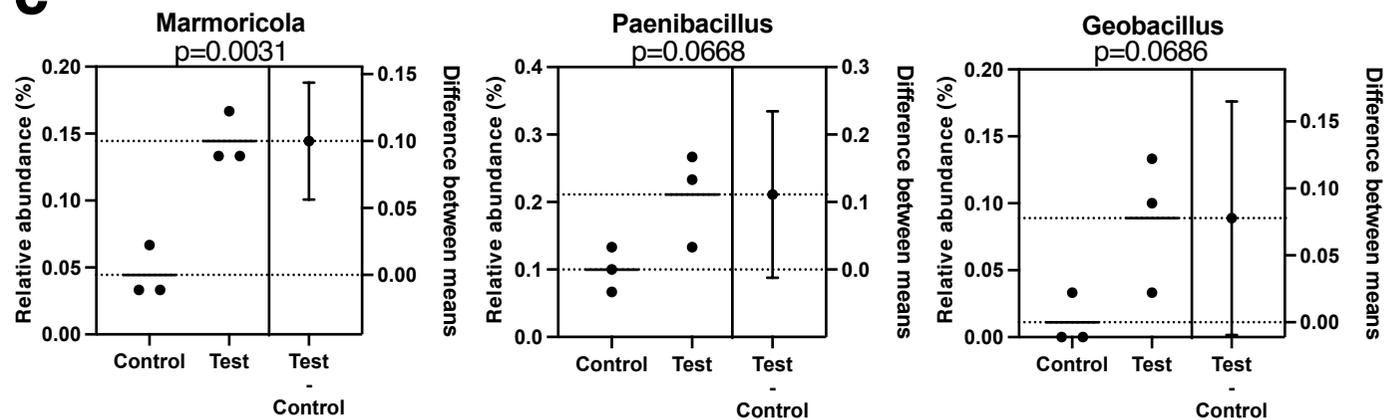

**Fig. 3**

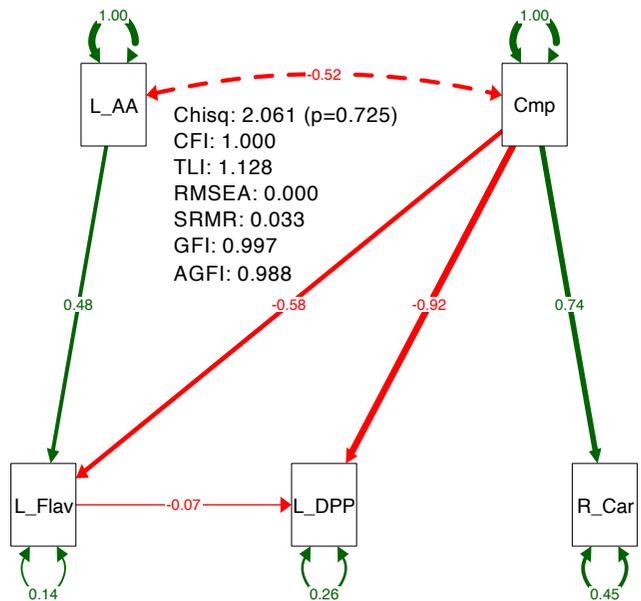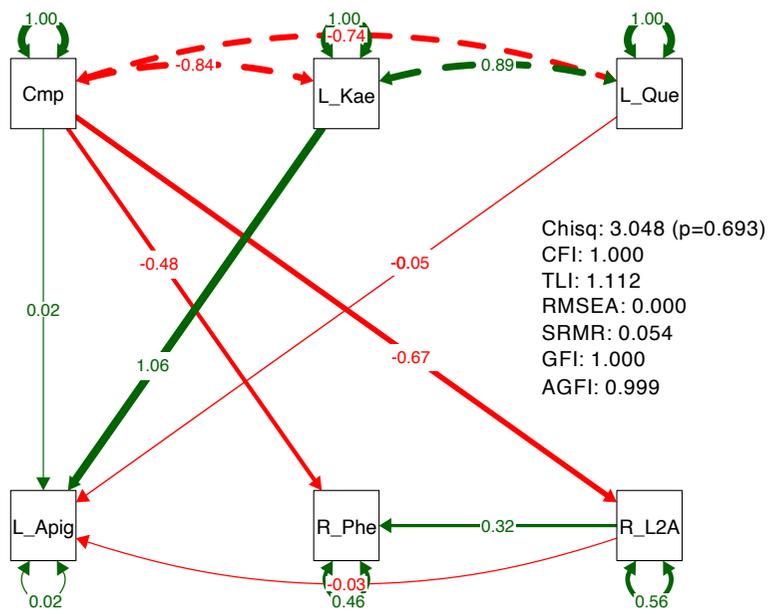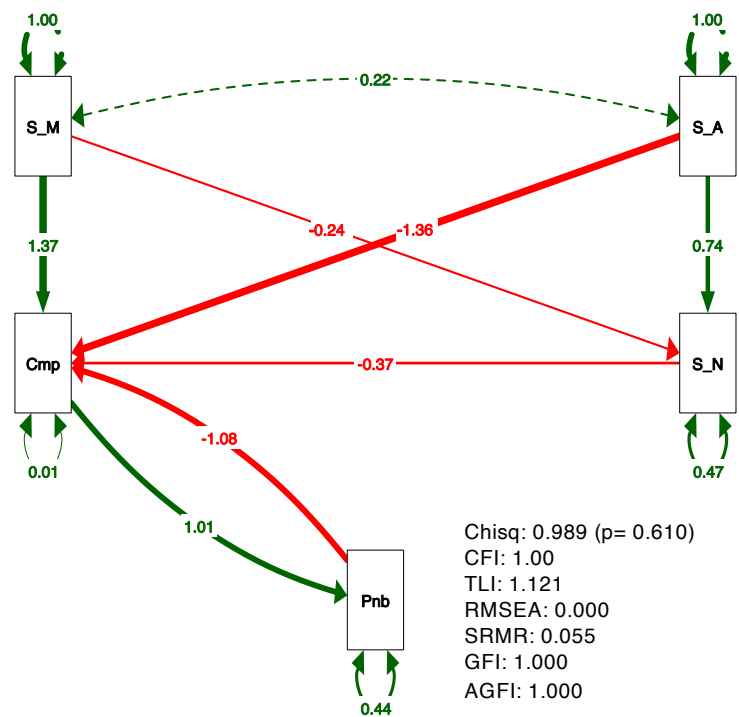

**Fig. 4**

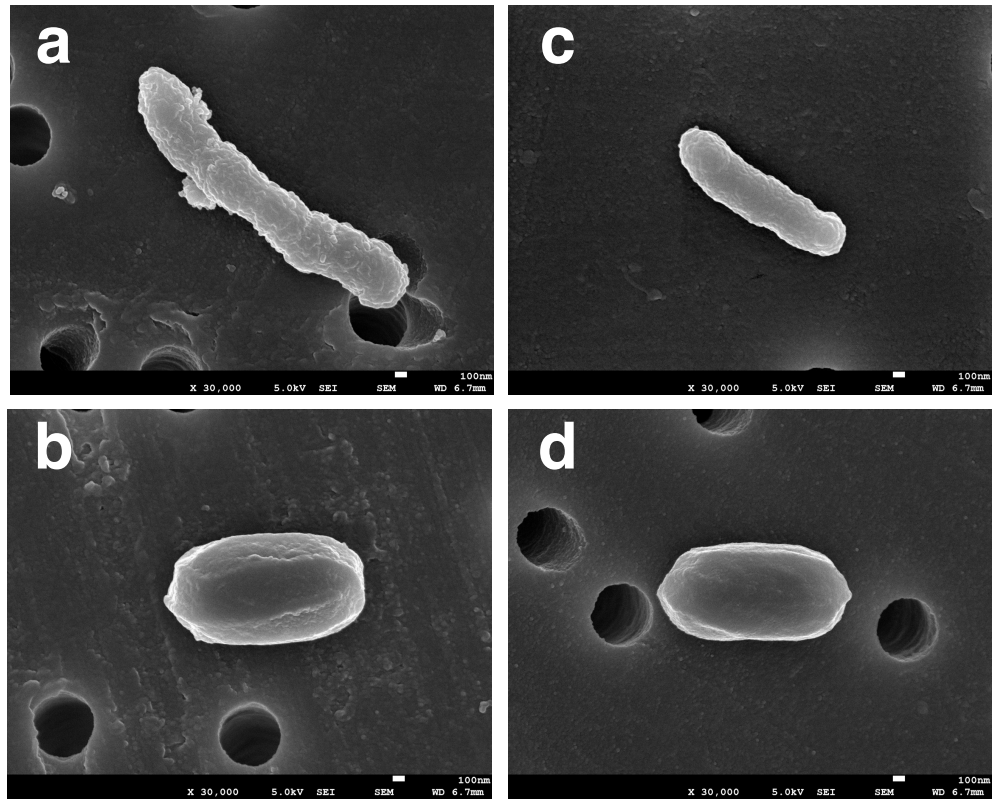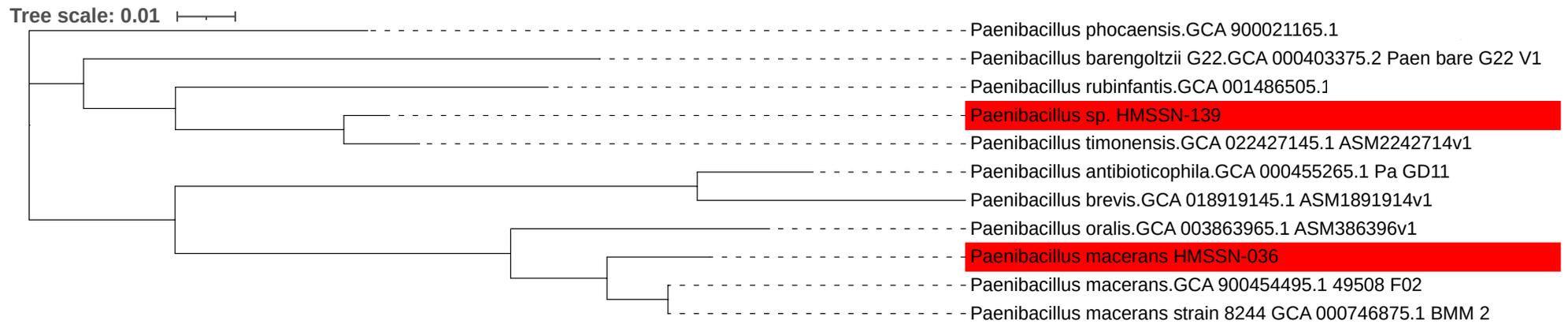

**Fig. 5**

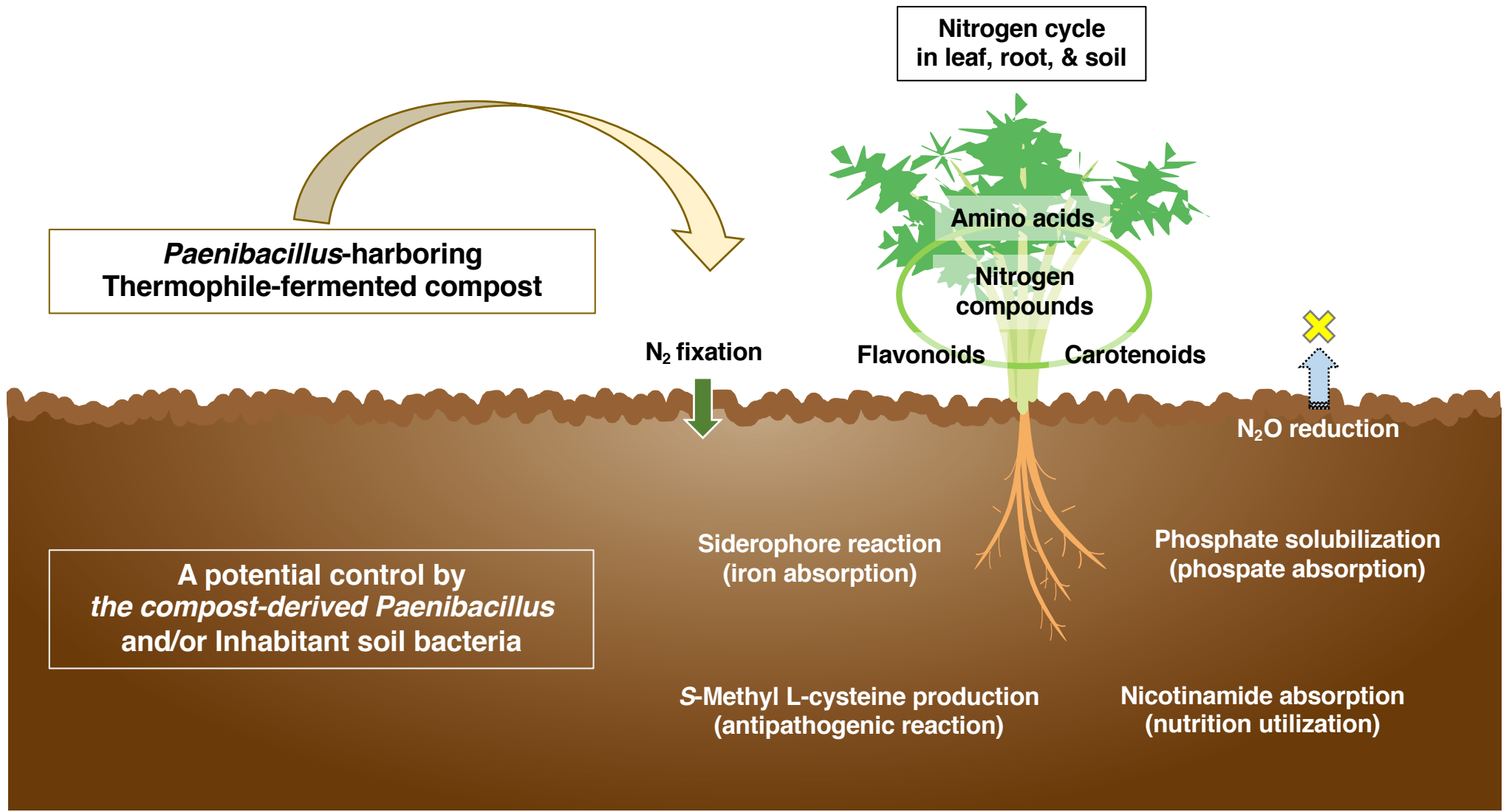

Fig. 6

## Table 1

| Function | Isolated strain names | |
|---|---|---|
| | *Paenibacillus macerans* HMSSN-036 | *Paenibacillus* sp. HMSSN-139 |
| Nitrogen fixation | nitrogen fixation protein NifB | nitrogen fixation protein NifB |
| | nitrogenase molybdenum-cofactor synthesis protein NifE | nitrogenase molybdenum-cofactor synthesis protein NifE |
| | nitrogenase iron protein NifH | nitrogenase iron protein NifH |
| | nitrogen fixation protein NifK | - |
| | nitrogenase molybdenum-iron protein NifN | nitrogenase molybdenum-iron protein NifN |
| | nitrogen fixation protein NifU | nitrogen fixation protein NifU |
| | nitrogen fixation protein NifX | nitrogen fixation protein NifX |
| Nitrogen cycle cofactor | nitrite transporter NirC | nitrite transporter NirC |
| Auxin production | auxin efflux carrier | auxin efflux carrier |
| | indole-3-glycerol phosphate synthase | indole-3-glycerol phosphate synthase |
| | auxin-induced protein | - |
| Phosphate solubilization | inorganic phosphate transporter, PiT family | inorganic phosphate transporter, PiT family |
| Siderophore reaction | Fur family transcriptional regulator, ferric uptake regulator | Fur family transcriptional regulator, ferric uptake regulator |

# Supplementary Information

## An agroecological structure model of compost-soil-plant interactions for sustainable organic farming

Miyamoto*[1,2,3,4], Katsumi Shigeta[5], Wataru Suda[2], Yasunori Ichihashi[6], Naoto Nihei[7], Makiko Matsuura[1,3], Arisa Tsuboi[4], Naoki Tominaga[5], Masahiko Aono[5], Muneo Sato[8], Shunya Taguchi[9], Teruno Nakaguma[1,3,4], Naoko Tsuji[3], Chitose Ishii[2,3], Teruo Matsushita[3,4], Chie Shindo[2], Toshiaki Ito[10], Tamotsu Kato[2], Atsushi Kurotani[8,11], Hideaki Shima[8], Shigeharu Moriya[12], Satoshi Wada[12], Sankichi Horiuchi[13], Takashi Satoh[14], Kenichi Mori[1,3,4], Takumi Nishiuchi[15], Hisashi Miyamoto[3,16], Hiroaki Kodama[1], Masahira Hattori[2,17,18], Hiroshi Ohno[2], Jun Kikuchi*[8], Masami Yokota Hirai*[8]

*Affiliations:*
1. Graduate School of Horticulture, Chiba University, Matsudo, Chiba 271-8501, Japan
2. RIKEN Center for Integrative Medical Science, Yokohama, Kanagawa 230-0045, Japan
3. Sermas Co., Ltd., Ichikawa, Chiba 272-0033, Japan
4. Japan Eco-science (Nikkan Kagaku) Co. Ltd., Chiba, Chiba 260-0034, Japan
5. Takii Seed Co.Ltd., Konan, Shiga 520-3231, Japan
6. RIKEN BioResource Research Center, Tsukuba, Ibaraki 305-0074, Japan
7. Faculty of Food and Agricultural Sciences, Fukushima University, Fukushima, Fukushima 960-1296, Japan
8. RIKEN Center for Sustainable Resource Science, Yokohama, Kanagawa 230-0045, Japan
9. Center for Frontier Medical Engineering, Chiba University, Chiba, Chiba 263-8522, Japan
10. Keiyo Gas Energy Solution Co. Ltd., Ichikawa, Chiba 272-0033, Japan
11. Research Center for Agricultural Information Technology, National Agriculture and Food Research Organization, Tsukuba, Ibaraki, Japan, 305-0856
12. RIKEN, Center for Advanced Photonics, Wako, Saitama, Japan, 351-0198
13. Division of Gastroenterology and Hepatology, The Jikei University School of Medicine, Kashiwa Hospital, Chiba, Japan.
14. Division of Hematology, Kitasato University School of Allied Health Sciences, Sagamihara, Kanagawa 252-0329, Japan
15. Division of Integrated Omics research, Bioscience Core Facility, Research Center for Experimental Modeling of Human Disease, Kanazawa University, Kanazawa, Ishikawa, 920-8640, Japan
16. Miroku Co.Ltd., Kitsuki, Oita 873-0021, Japan
17. School of Advanced Science and Engineering, Waseda University, Tokyo169-8555, Japan
18. School of Agricultural and Life Sciences, The University of Tokyo, Bunkyo, Tokyo 113-8657, Japan

\* Cocorrespondence:

Hirokuni Miyamoto Ph.D., hirokuni.miyamoto@riken.jp; h-miyamoto@faculty.chiba-u.jp

Jun Kikuchi Ph.D., jun.kikuchi@riken.jp

Masami Yokota Hirai Ph.D., masami.hirai@riken.jp

# Figure contents

# Table contents

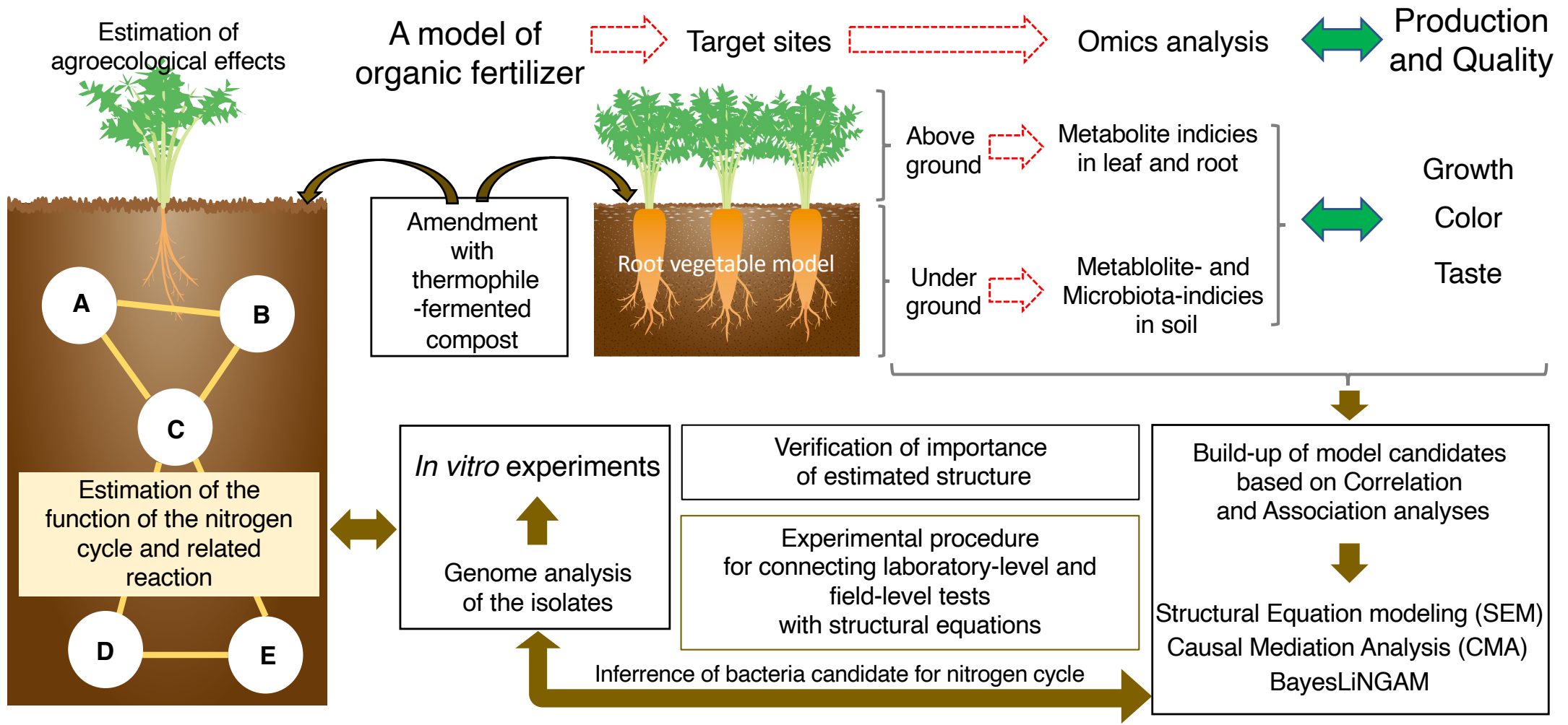

**Fig. S1**

Experimental scheme in this study. As a model of organic fertilizer, thermophile fermented compost was used, and as a crop model, carrots, which are easy to analyze both above and below the ground, were targeted. The leaves and roots, as well as soil metabolites, were analyzed, and the soil flora was analyzed to evaluate its relationships with growth, color, and taste as agricultural quality indices. These relationships were evaluated by correlation and association analysis, hypotheses were made, and a structural equation modeling (SEM) was constructed. Since one of the bacterial candidates for nitrogen cycle was selected from the structural equation for the soil, the candidate bacteria were isolated, and their genomes were analyzed. Nitrogen fixation and suppression of nitrous oxide ($N_2O$) generation was also analyzed at the laboratory level. An analysis of the suppression of $N_2O$ generation, which is difficult to perform in the field, was conducted in the laboratory. This study was performed using a procedure that connects laboratory-level and field-level tests with SEM to support experiments that cannot be conducted. In addition, causal mediation analysis (CMA) and BayesLiNGAM verified the importance as a group of optimal model candidates estimated by SEM. Based on these observations, a model for the agroecological effects of compost was estimated.

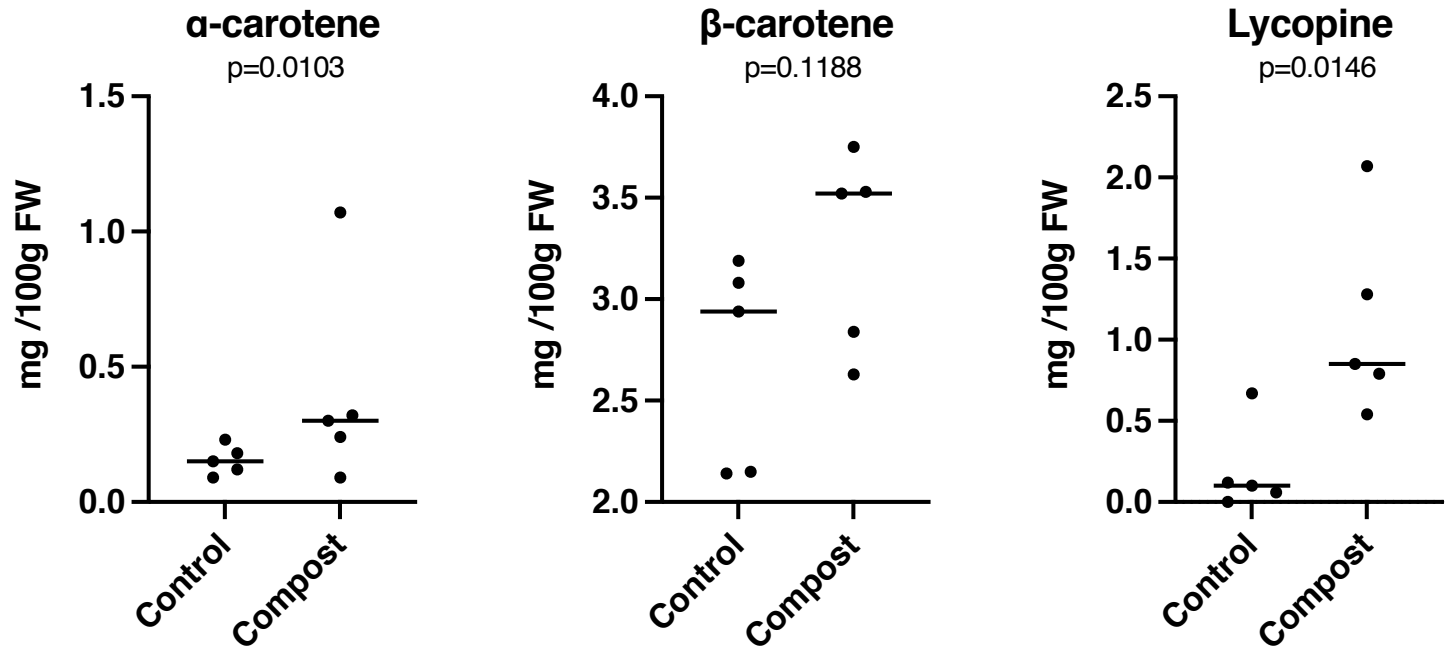

**Fig. S2**
The contents of primary carotenoids on the roots of carrots on the leaves and roots of carrots. The content of α-carotene, β-carotene, and lycopene (n=5) was shown. The "Control" and "Compost" show the data under normal conditions (control group) and compost-amended conditions (compost group), respectively.

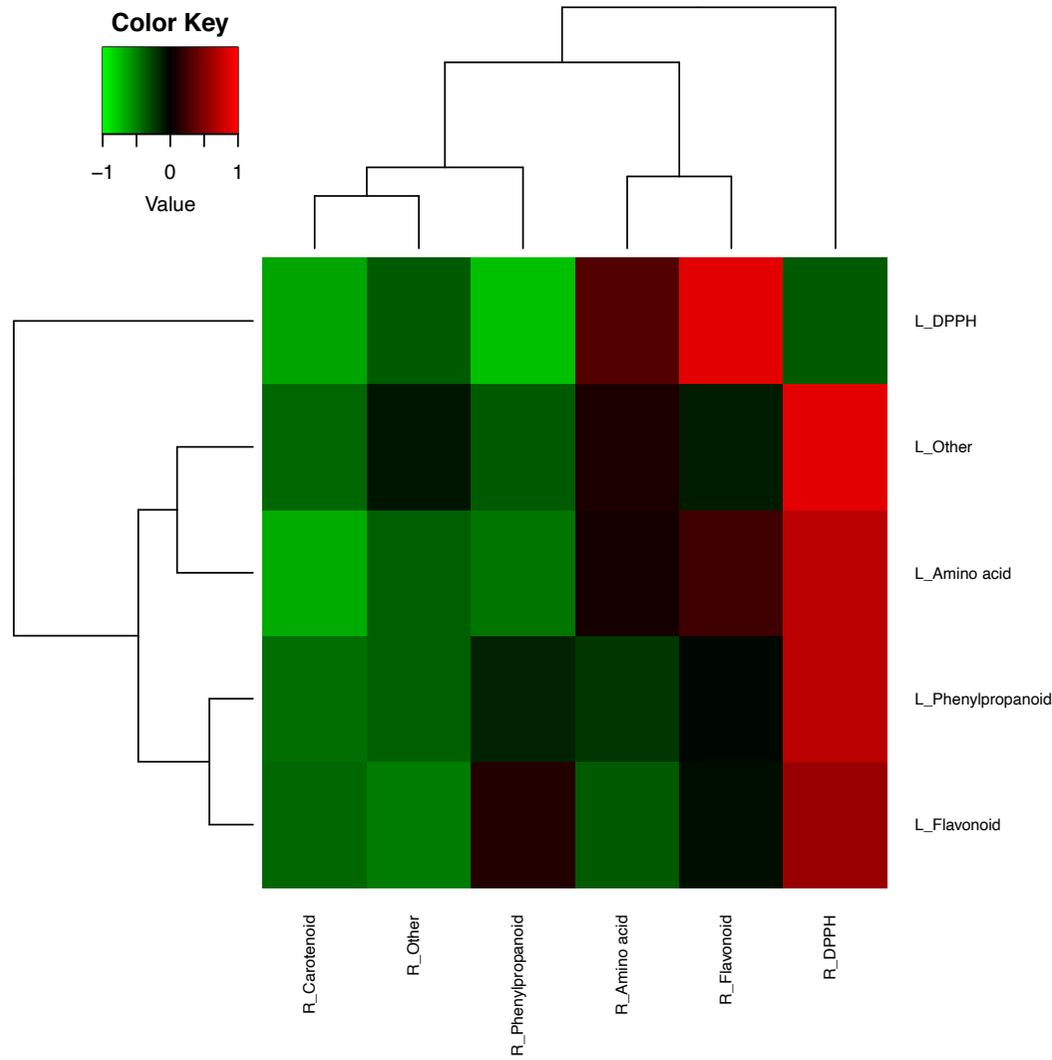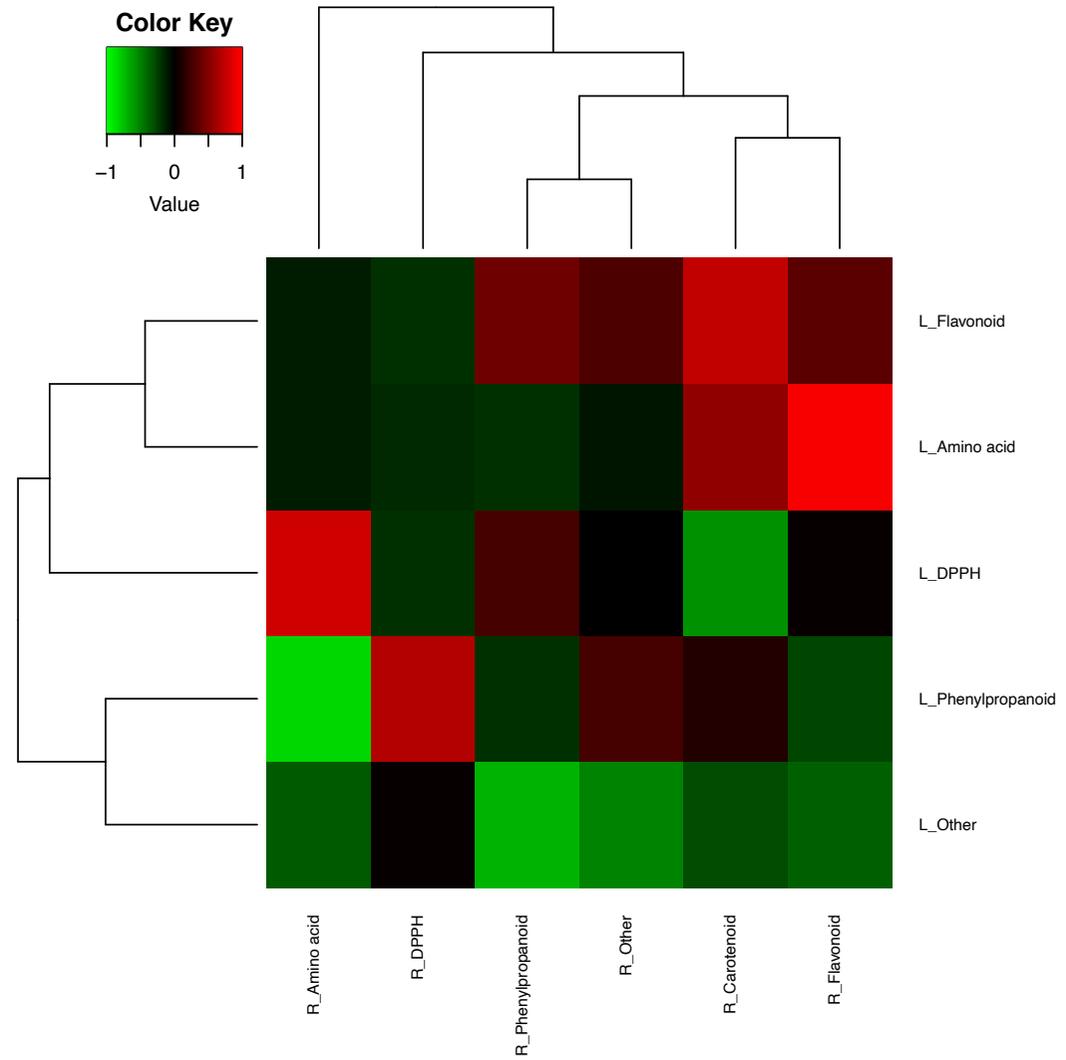

**Fig. S3**
Heatmaps of the correlations between the amino acid, carotenoid, flavonoid, and DPPH activity in the leaves and roots. The heatmaps based on data in (a) the control group and (b) the test group were shown.

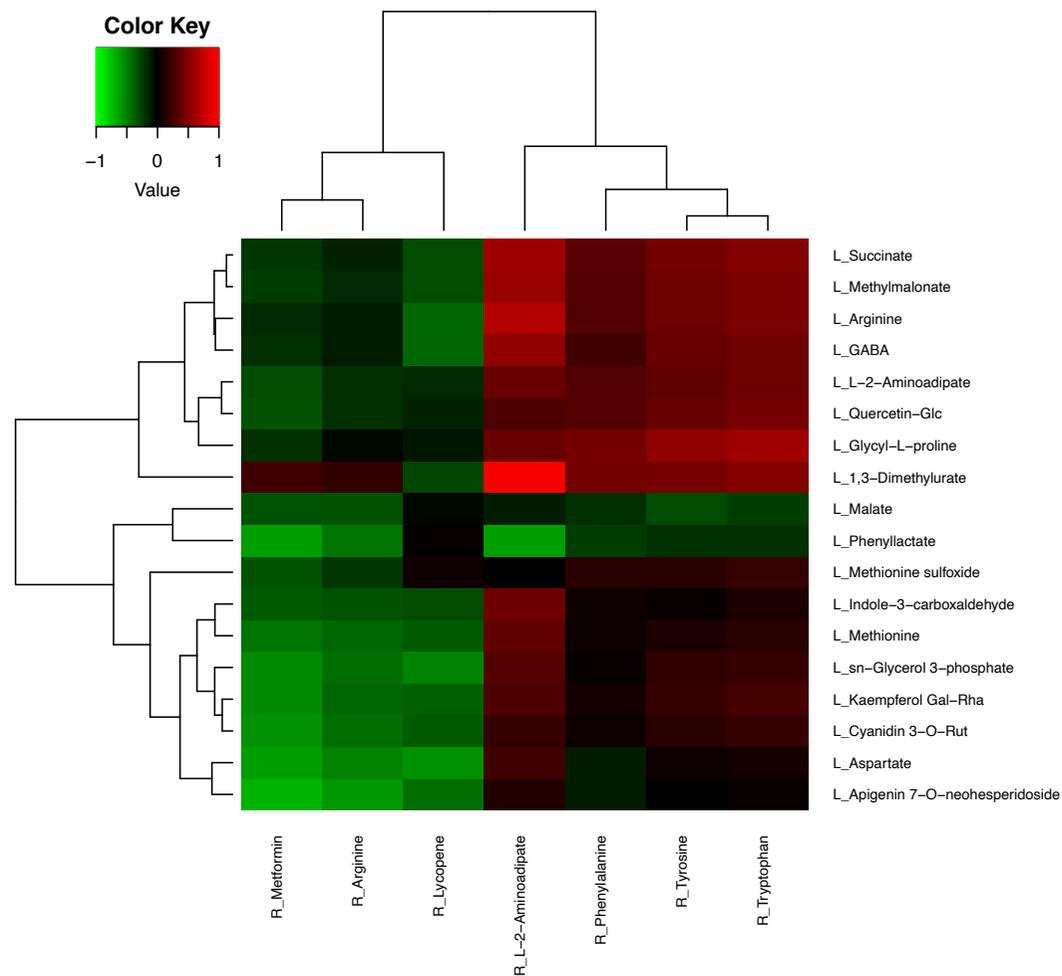 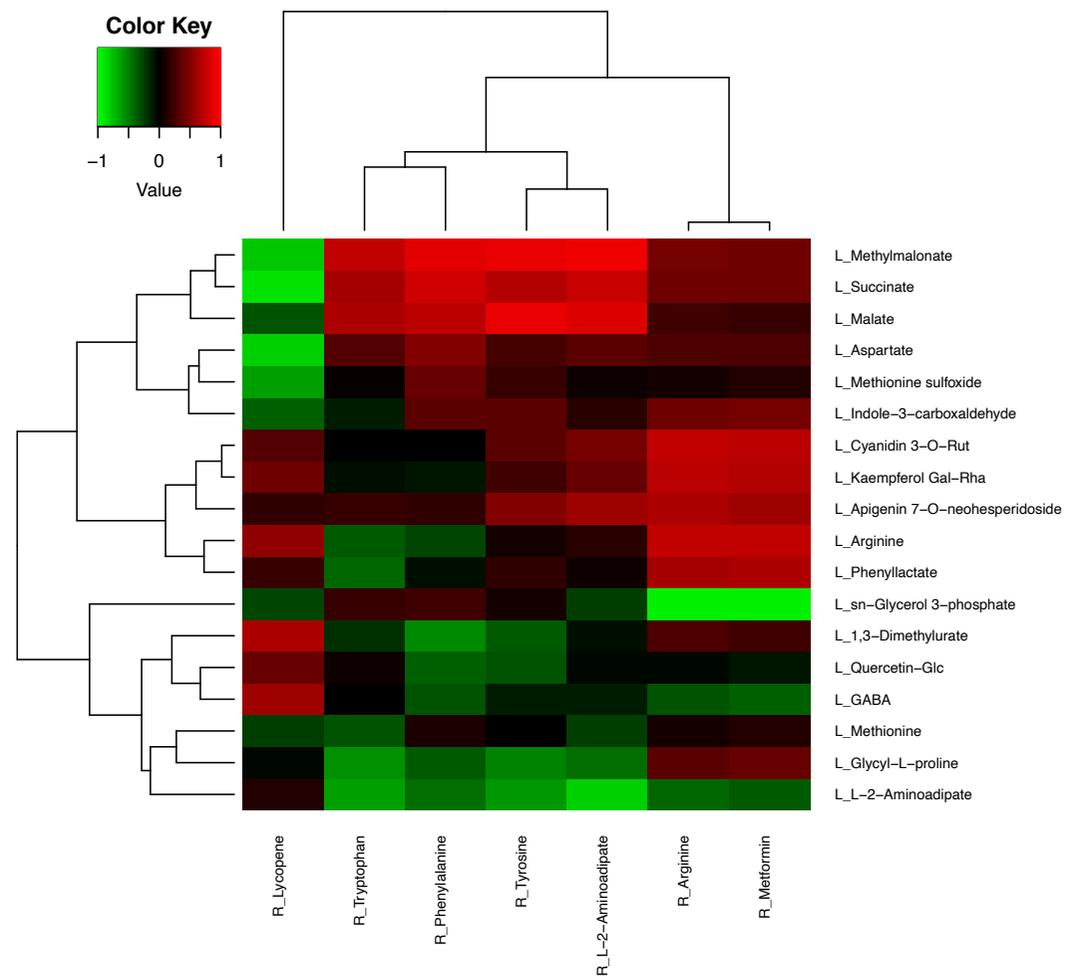

**Fig. S4**
Heatmaps of the correlations between metabolite candidates in the leaves and roots shown in Fig. 2. The heatmaps based on the data in (a) the control group and (b) the test group were shown.

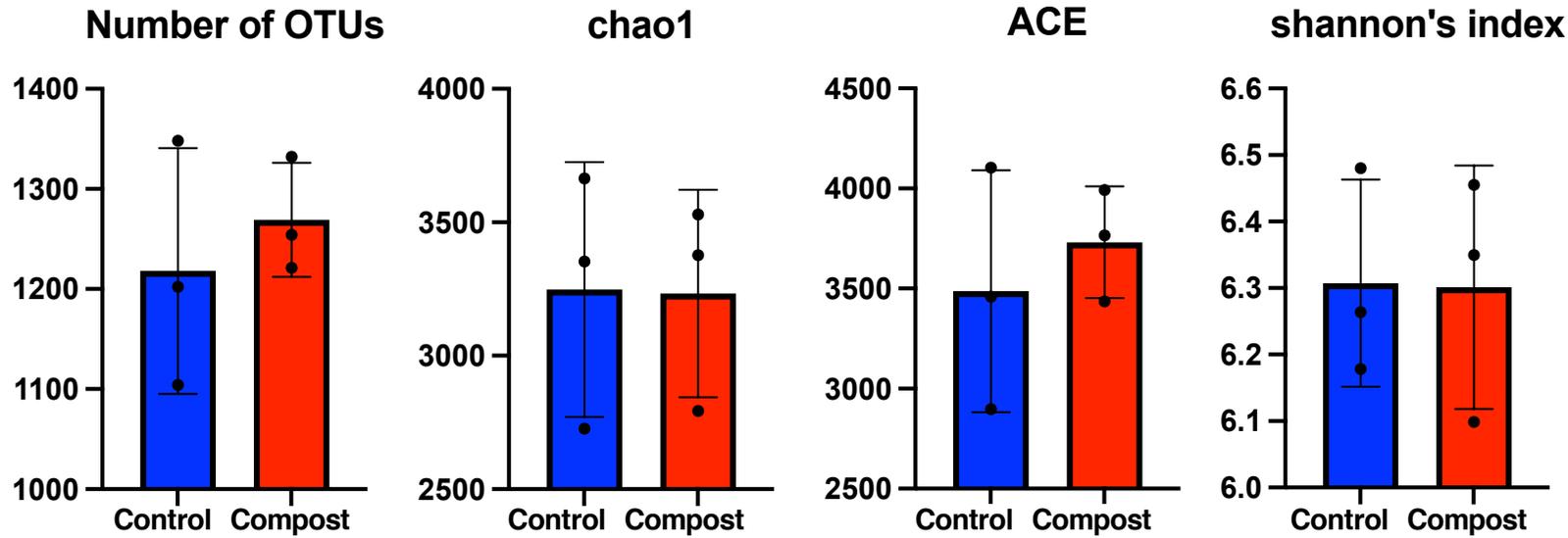
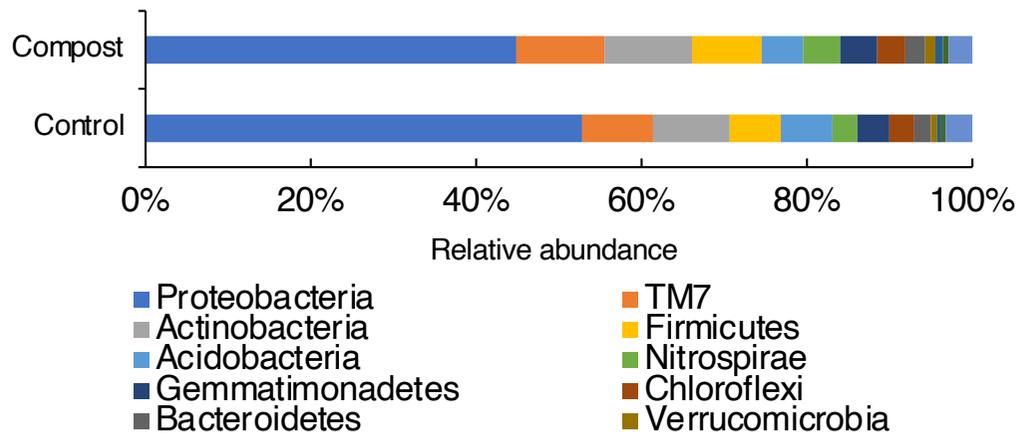
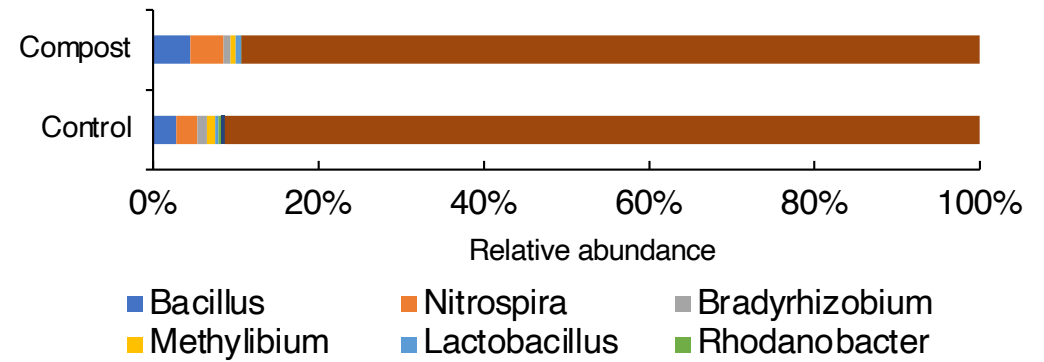

**Fig. S5**

Bacterial diversity in the soil after cultivation of carrot. (a) OTU numbers and Chao1, ACE, and Shannon indices representing α-diversity under normal conditions (Control) and compost-amended conditions (Compost). The bacterial population in the soil of the control and test groups showing as relative abundances of the (b) phyla and (c) genera (>1% as maximum of the bactereiral population).

a
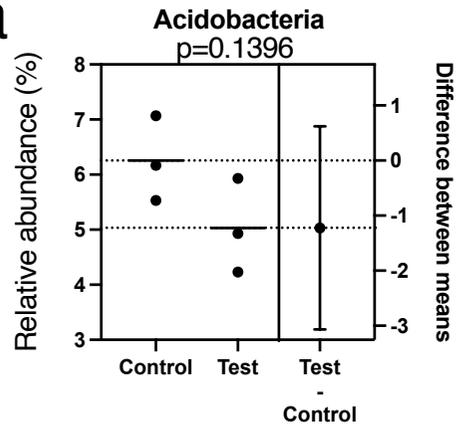
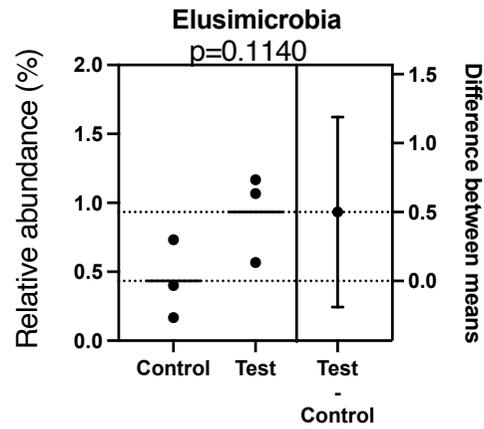

**Fig. S6**
The estimation plot of the bacterial population in the soil of the control and test groups. Relative abundances of the (a) phyla and (b) genera (0.1 < p <0.2; >0.1% as maximum of bacterial population).

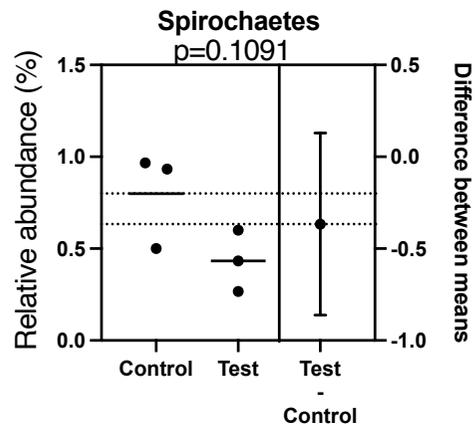
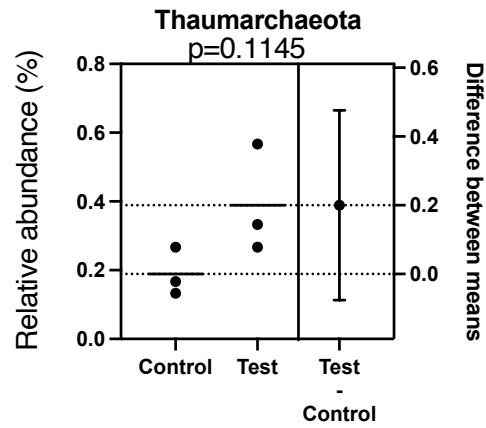
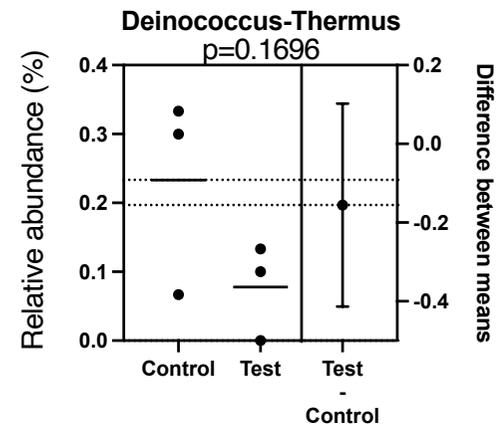

b
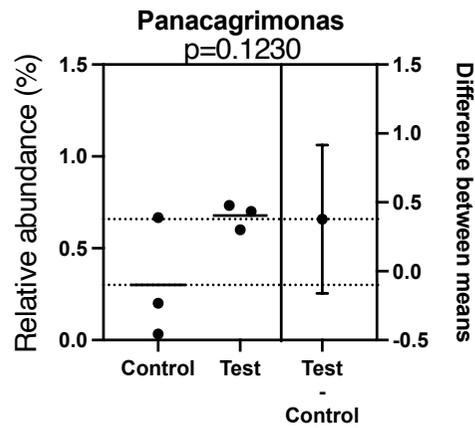
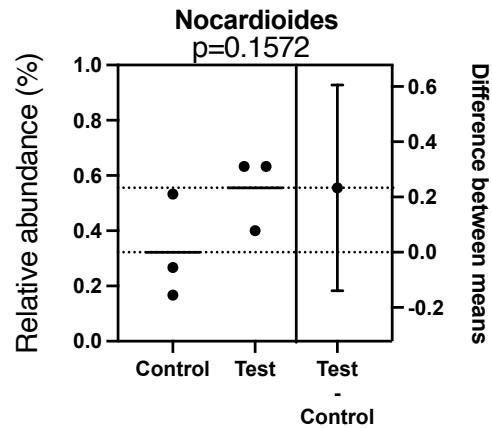
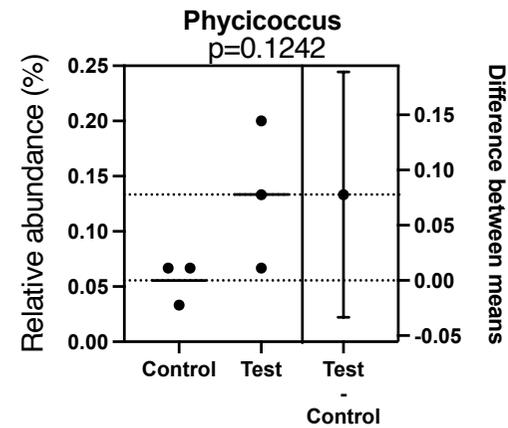

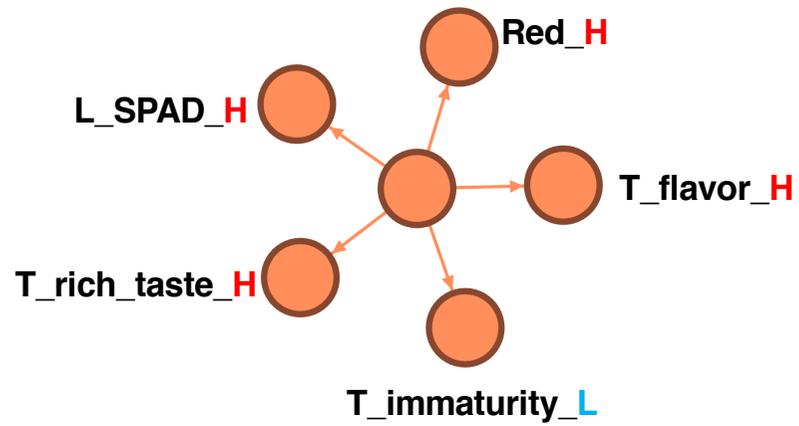
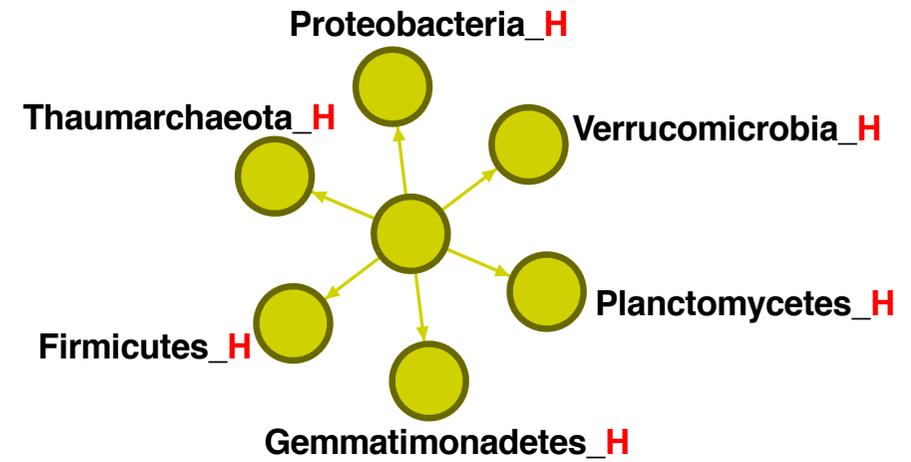
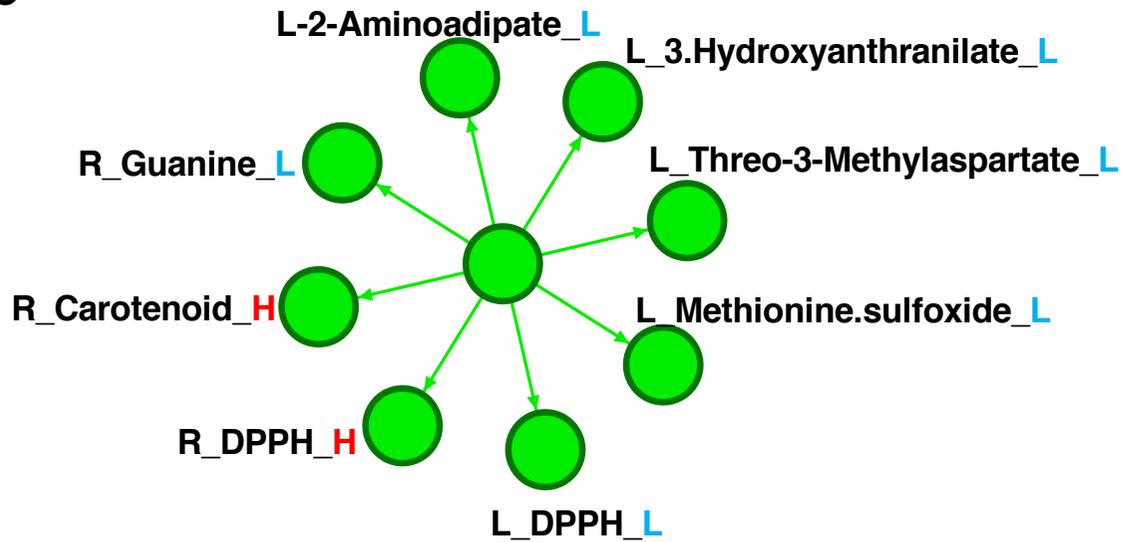
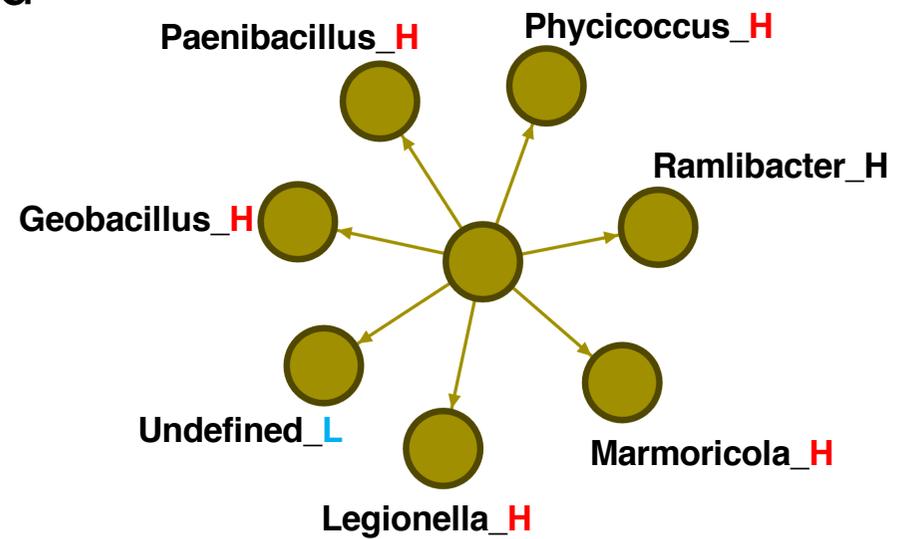

**Fig. S7**
Association network in (a) appearance and taste, (b) leaves and root metabolites, (c) soil bacterial phyla and (d) genera (lift values > 2.0).

a

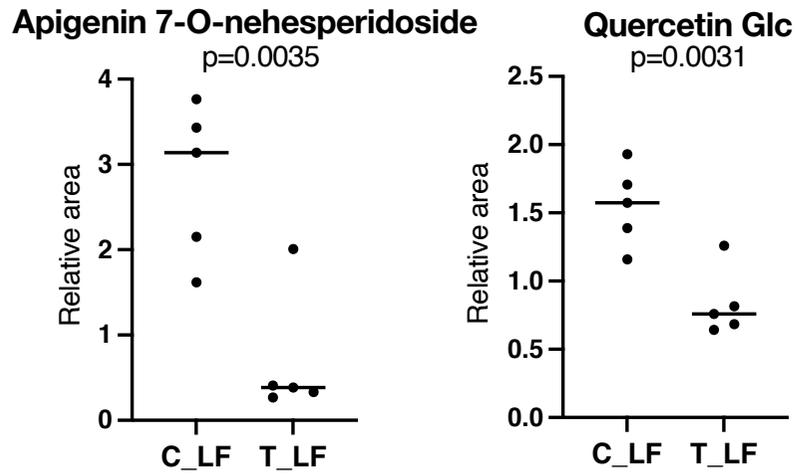
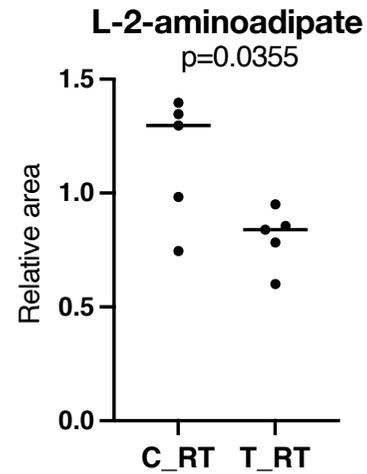

b

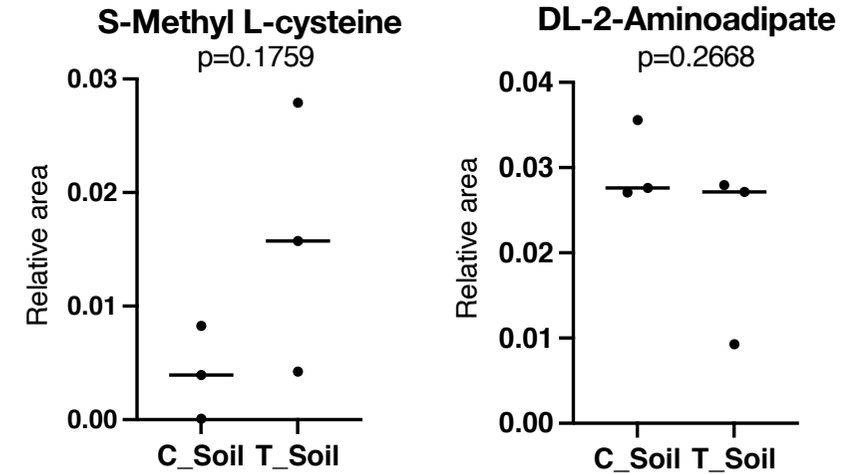

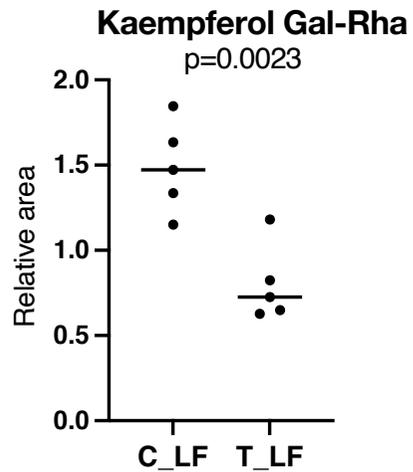
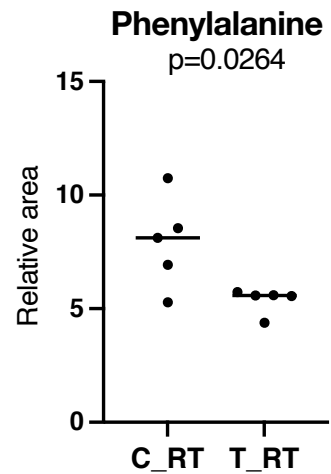
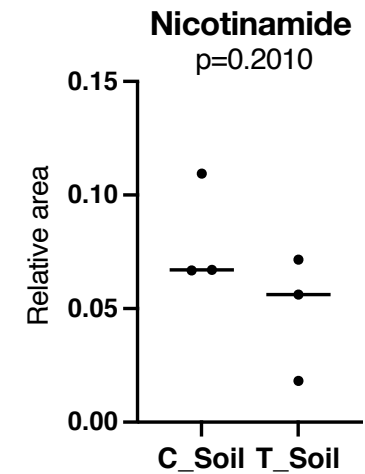

**Fig. S8**

Degree of detection in metabolite candidates used for optimal structural equations. (a) The relative area of metabolite candidates in Fig. 4c. (b) The relative area of metabolite candidates in Fig 5a.

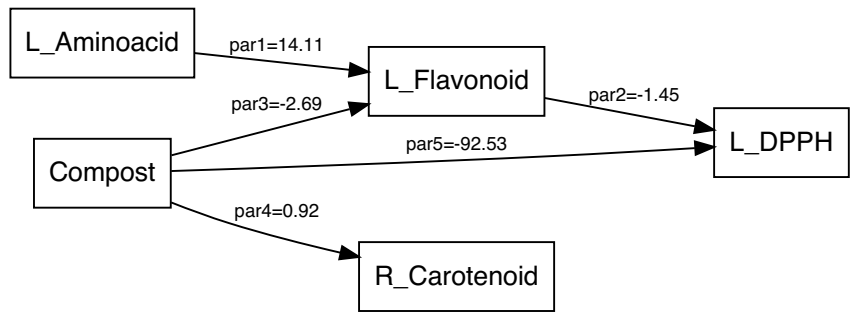
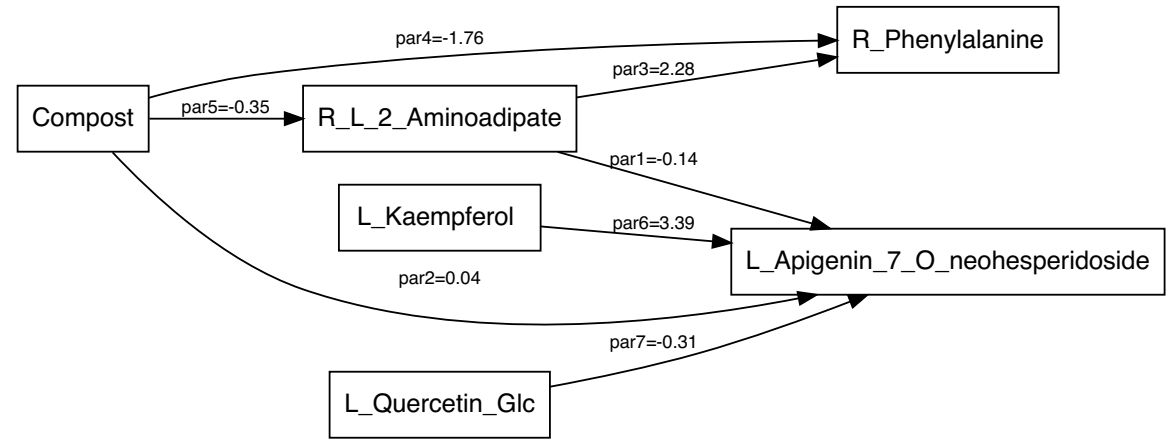
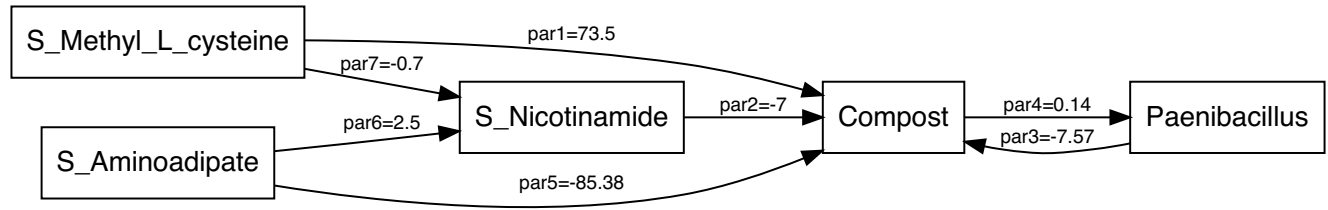

**Fig. S9**
The path of Fig. 4 calculated by the function sem. (a), (b), and (c) shows the path in Figs. 4a, 4b, and 5a, respectively. The value in the back of each "par=" shows the "startvalue" as the parameter (par) of the indicated each path. The number in back of each "par=" shows just a simple order of description.

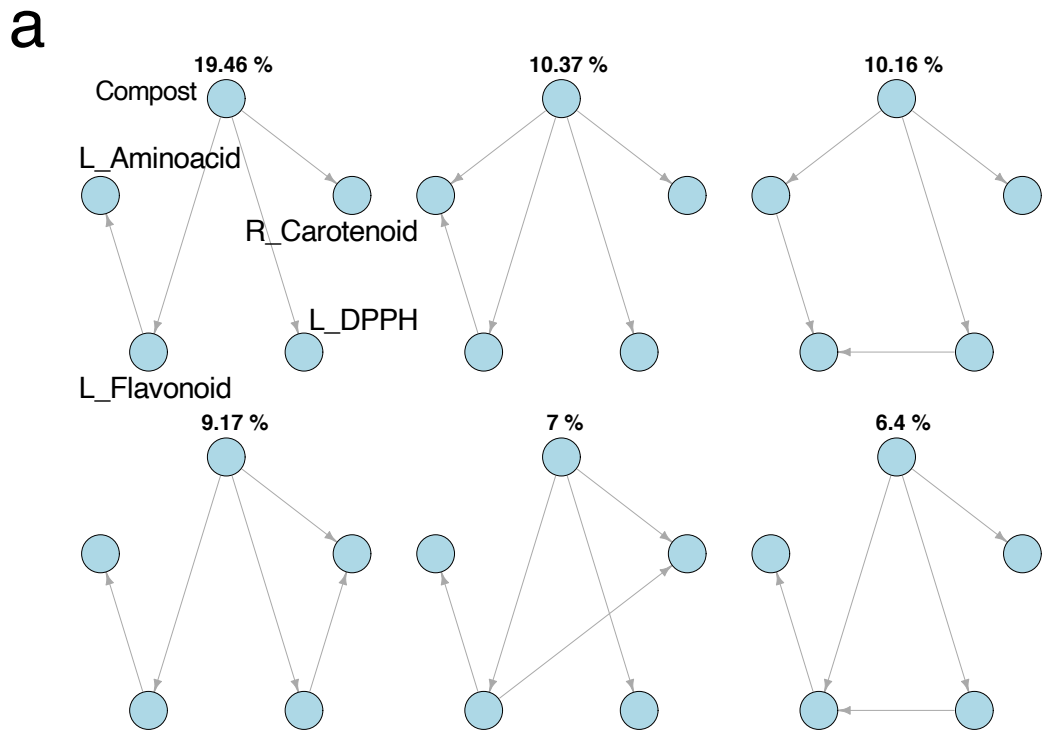
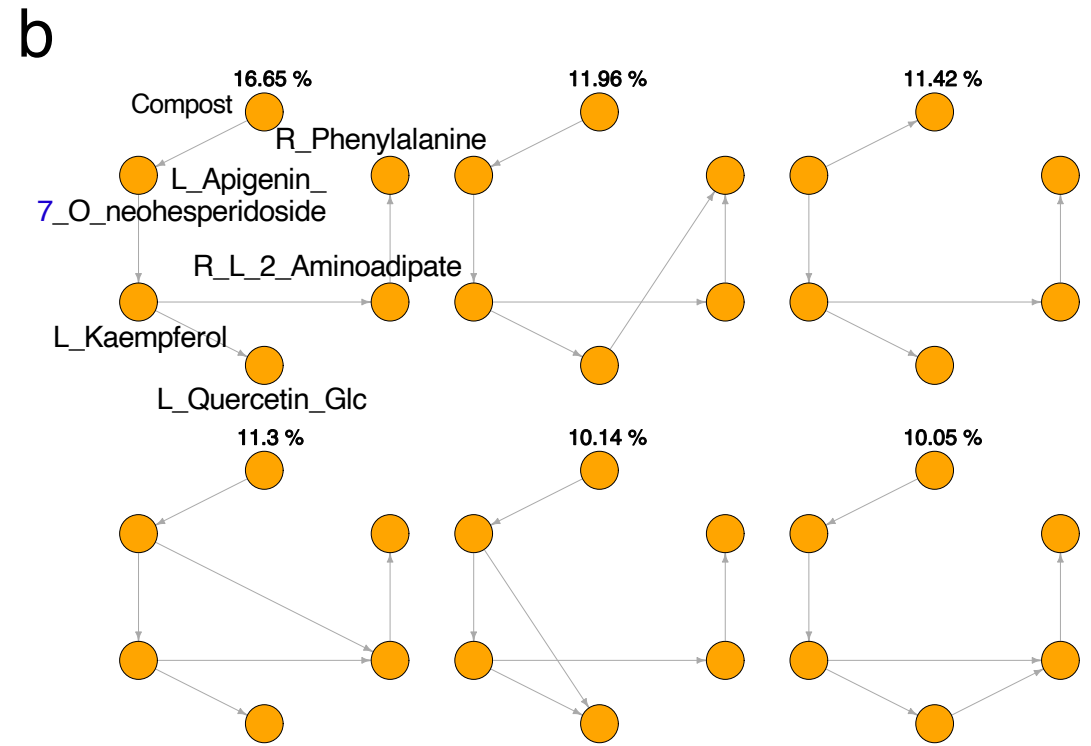
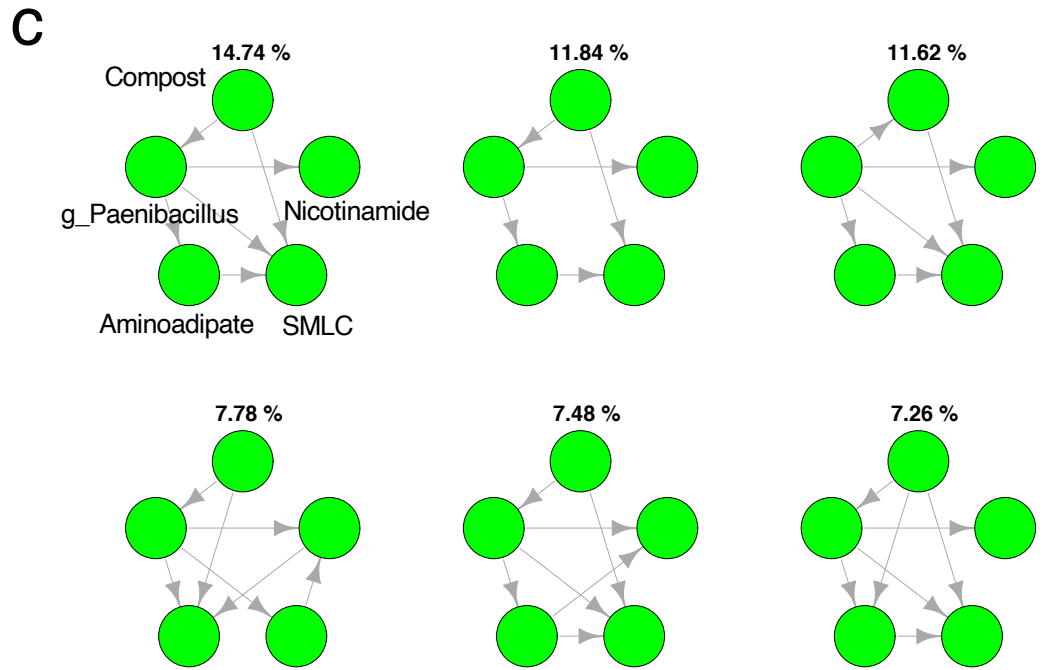

**Fig. S10**

The top six causal structural groups for SEM (Fig. 4) estimated by BayesLiNGAM. The causal groups for (a) Fig. 4a, (b) Fig. 4b, and (c) Fig.4c are shown. The direction of causality and its probability are shown by the arrows and percentages, respectively. Each component name is representatively listed in the position that indicates the percentage of the top (upper left side), respectively. The arrangement of the component names was also fixed within the other categories. The abbreviations in the table indicate the following: L_, leaf metabolites or activities; R_, root metabolites; g_, genus; SMLC, S-Methyl L-cysteine.

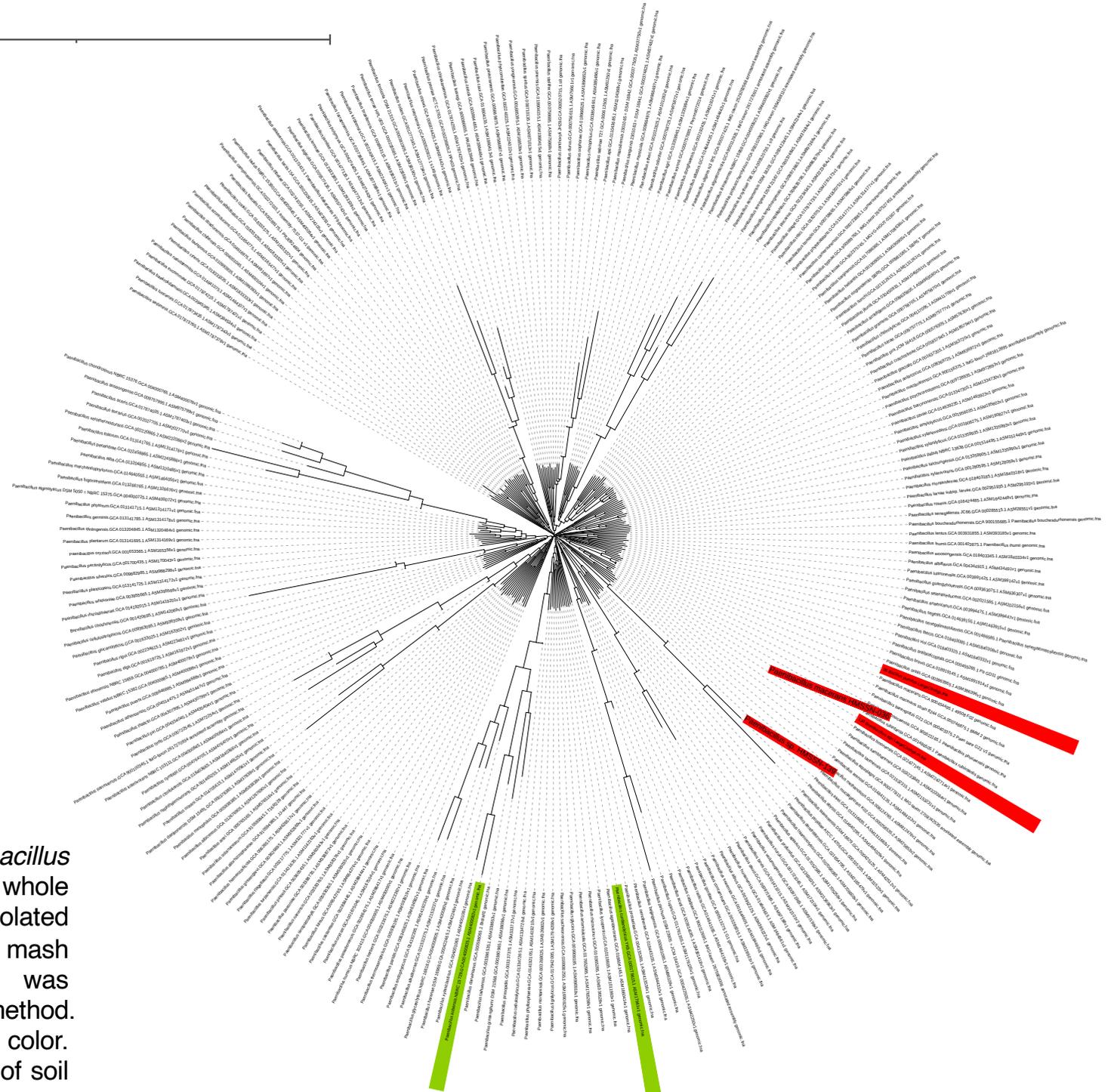

**Fig. S11**
Phylogenetic relationship among *Paenibacillus* strains isolated in this study. The whole genome phylogenetic analysis of the isolated strains was performed based on the mash distance. The phylogenetic tree was constructed using the neighbor-joining method. The isolated strains are marked in red color. Species detected by 16SrRNA analysis of soil are marked in green.

a

| strain | IAA conc. (μg/mL) | phosphate solubilization | siderophore reaction |
| --- | --- | --- | --- |
| #36 strain | 5.934 | positive | positive |
| #139 strain | 5.857 | positive | positive |

b

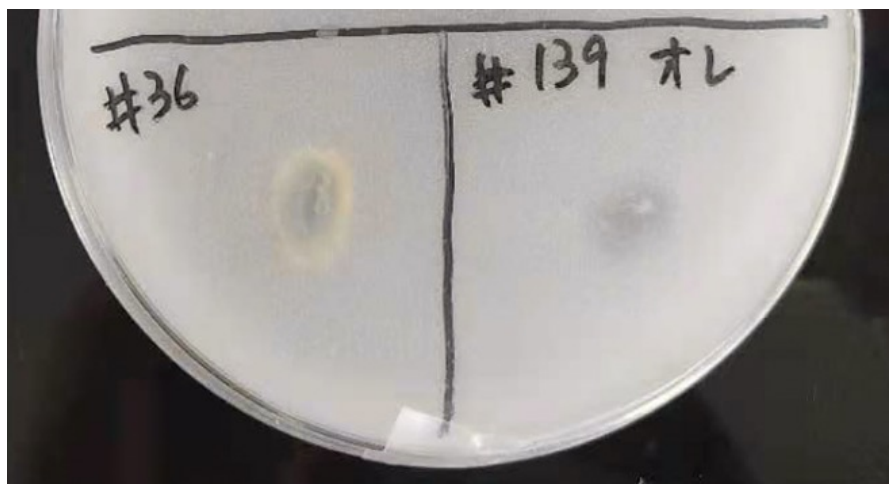

c

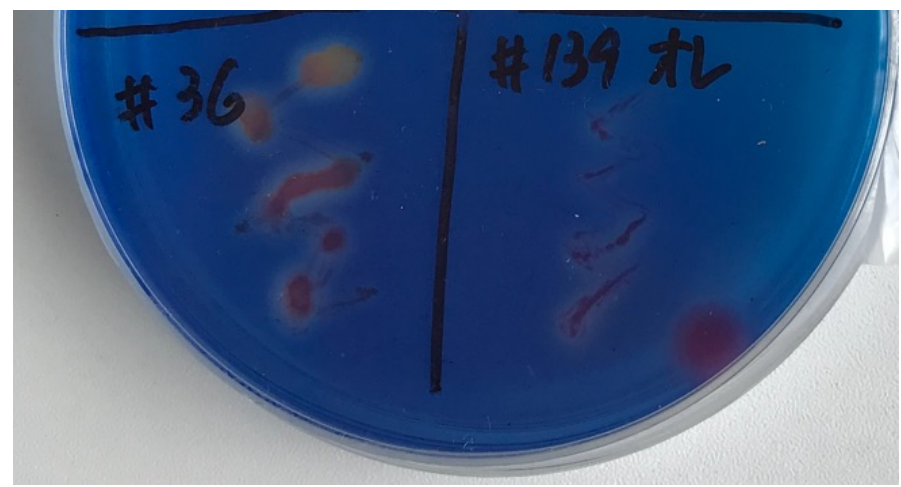

**Fig. S12**
Biological assay for isolated Paenibacillus strains. (a) Summarized data from the biological assay. The photos of the (b) phosphate solubilization and (c) siderophore reaction tests of selected strains are shown.

**a**

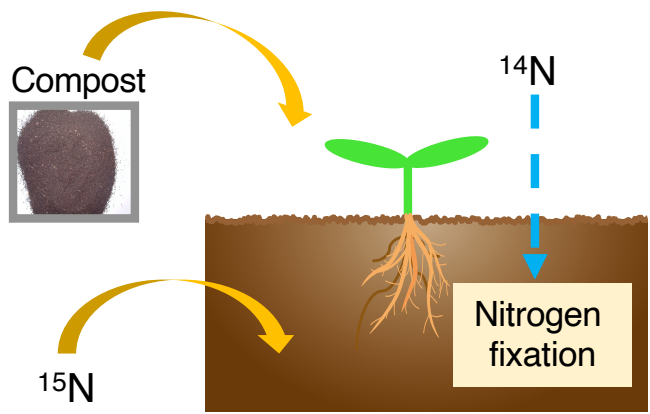

### Fig. S13

*In vitro* assay to evaluate (a)(b) nitrogen ($N_2$) fixation and (c)(d) nitrous oxide ($N_2O$) generation from from the soil. (a) Conceptual diagram of the test model with the stable isotope $^{14}N$ is shown. The experiments to evaluate nitrogen fixation were carried out in the test model. (b) Biological weight of plants in the stable isotope test. The ratios of the $^{14}N$ and the stable isotopes $^{15}N$ contained in freeze-dried crop and soil samples are shown. S-Compost, the group to which sterile compost was added; Compost, the group to which compost was added; and Compost (60 °C), the group to which high temperature treated compost was added immediately before the experiment. (c) The conceptual diagram of the test model is shown. After a container containing soil was inserted into Gas package I, it was joined to Gas package II with a cock. These sets were prepared with different soil conditions. The composition of fungi in the prepared preexperimental soil was also investigated. (d) shows the $N_2O$ concentration generated from the soil. The "Soil only" indicates the soil only group. PD indicates potato dextrose. The contents of $N_2O$ from "Soil" and "Soil + Compost" groups were markedly different ($p < 0.1$).

**b**

| Category | Plant | | | | Soil | |
|---|---|---|---|---|---|---|
| | FW | n | $\delta^{15}N$ (‰) | $^{15}N / ^{14}N$ | $\delta^{15}N$ (‰) | $^{15}N / ^{14}N$ |
| Soil only | 0.832 ± 0.134 | 3 | 4.9 | 0.051 | 4.16 | 0.043 |
| Soil + S-Compost | 1.419 ± 0.348 | 3 | 4.68 | 0.049 | 3.18 | 0.033 |
| Soil + Compost | 1.608 ± 0.407 | 3 | 3.92 | 0.041 | 2.74 | 0.028 |
| Soil + Compst (60°C) | 1.252 ± 0.161 | 3 | 3.95 | 0.041 | 3.09 | 0.032 |

**c**

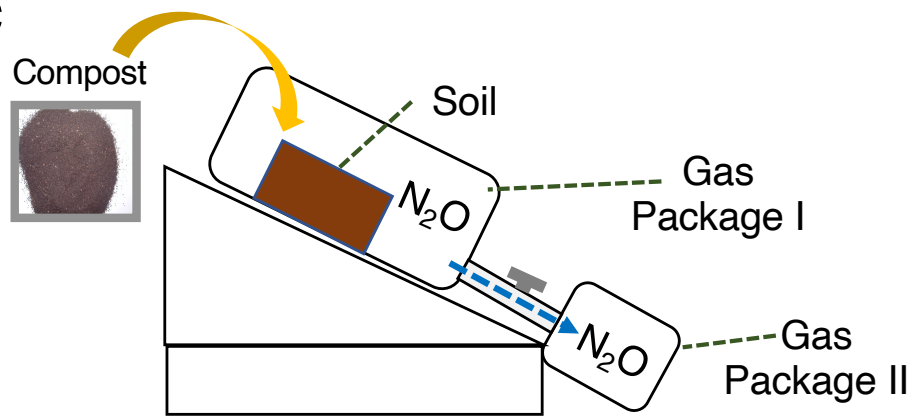

**d**

| Category | PDA | n | $N_2O$ (ppb) | p value |
|---|---|---|---|---|
| Soil (PDA-) | - | 1 | 120.31 | - |
| Soil only | + | 3 | 145.20 ± 10.37 | - |
| Soil + Compost | + | 3 | 119.59 ± 5.07 | 0.0684 |

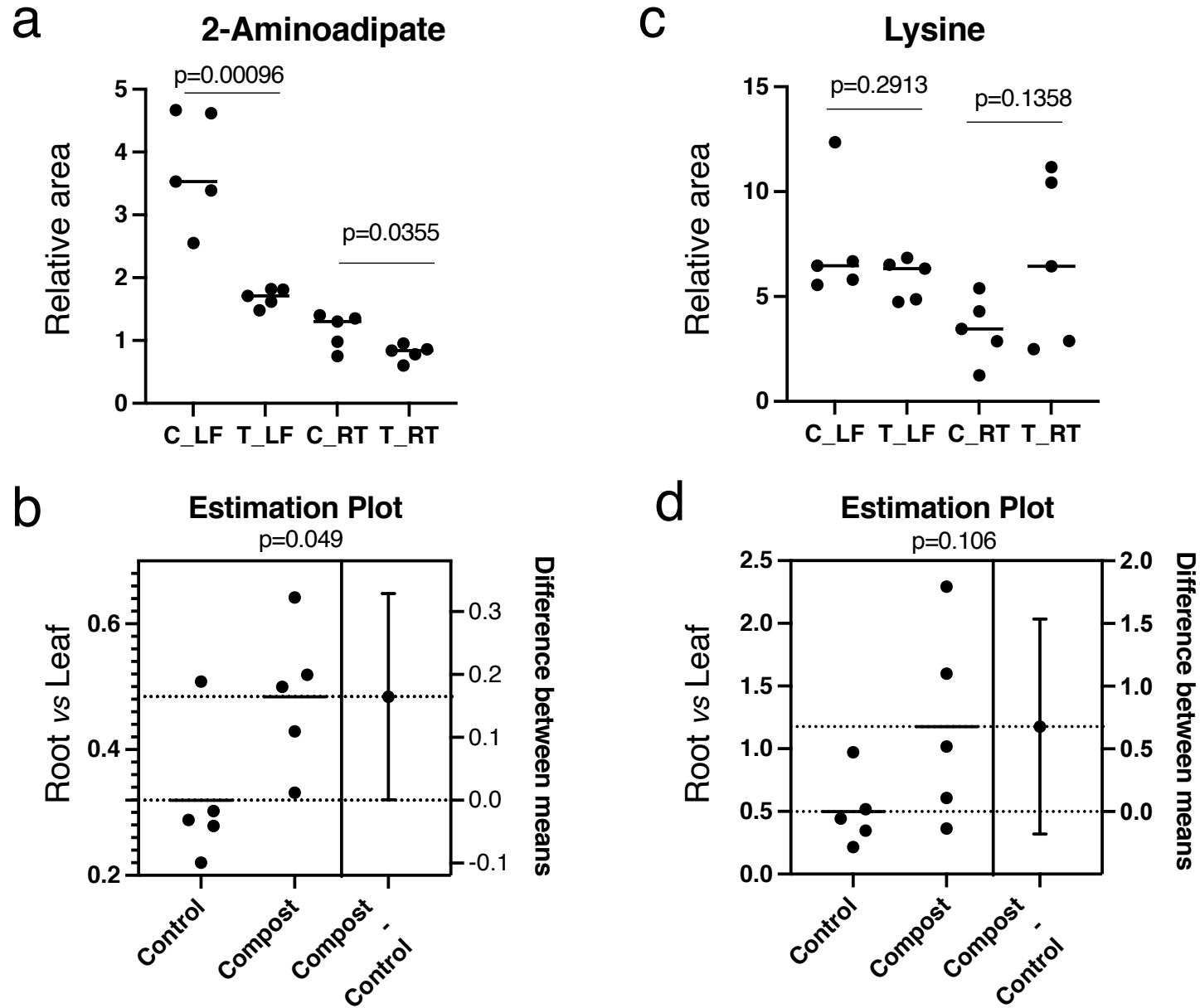

**Fig. S14**
Degree of detection in leaves and roots of 2-aminoadipate and lysine and their ratio in root per leaf. (a) shows the relative area of 2-aminoadipate standardized by the internal standard. (b) shows the ratio of 2-aminoadipate in the root per leaf. (c) shows the relative area of lysine standardized by the internal standard. (d) shows the ratio of lysine in root per leaf.

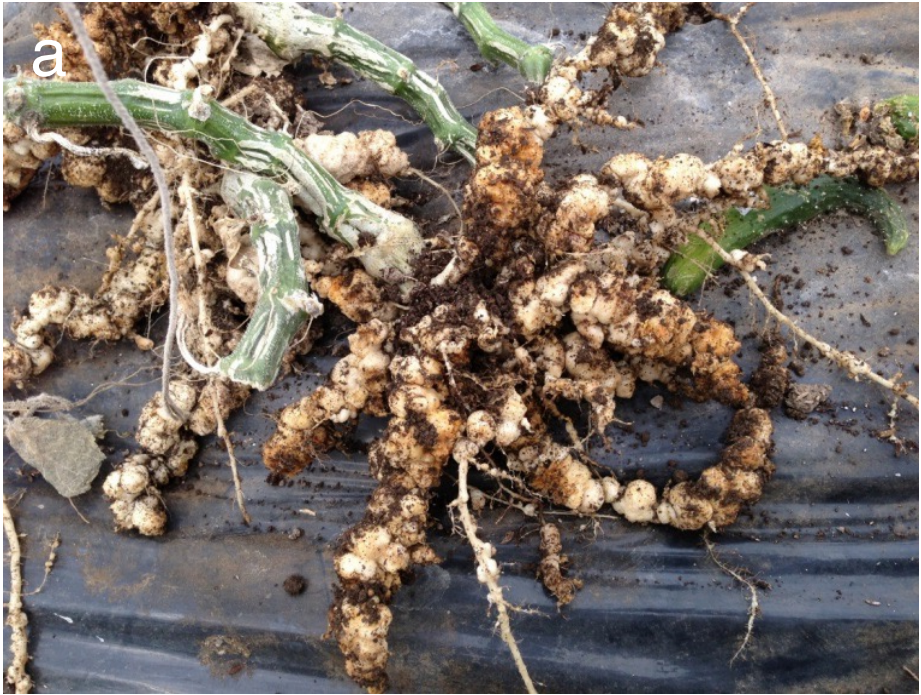
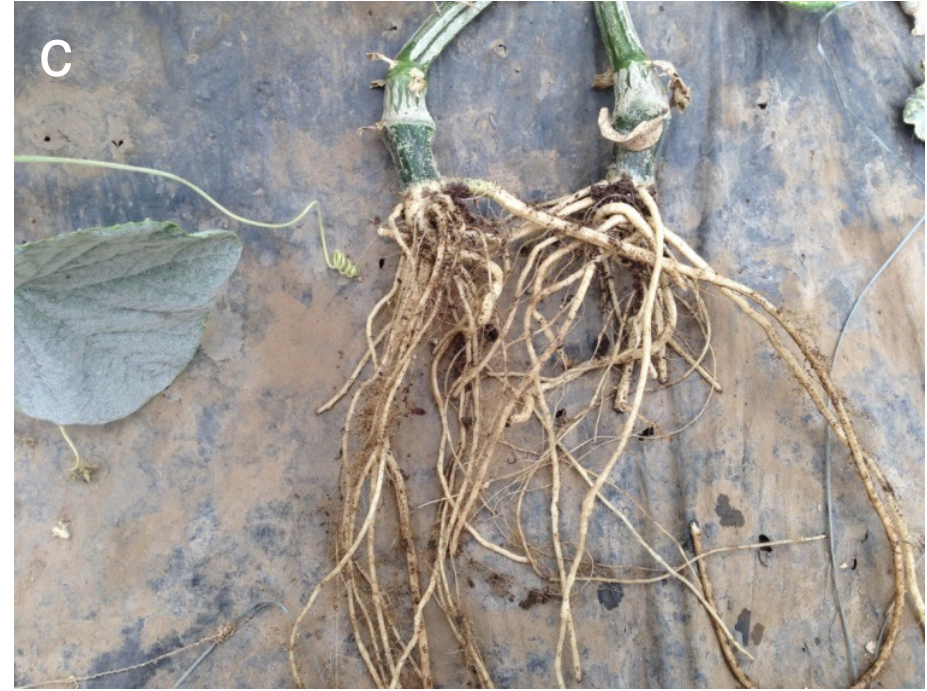
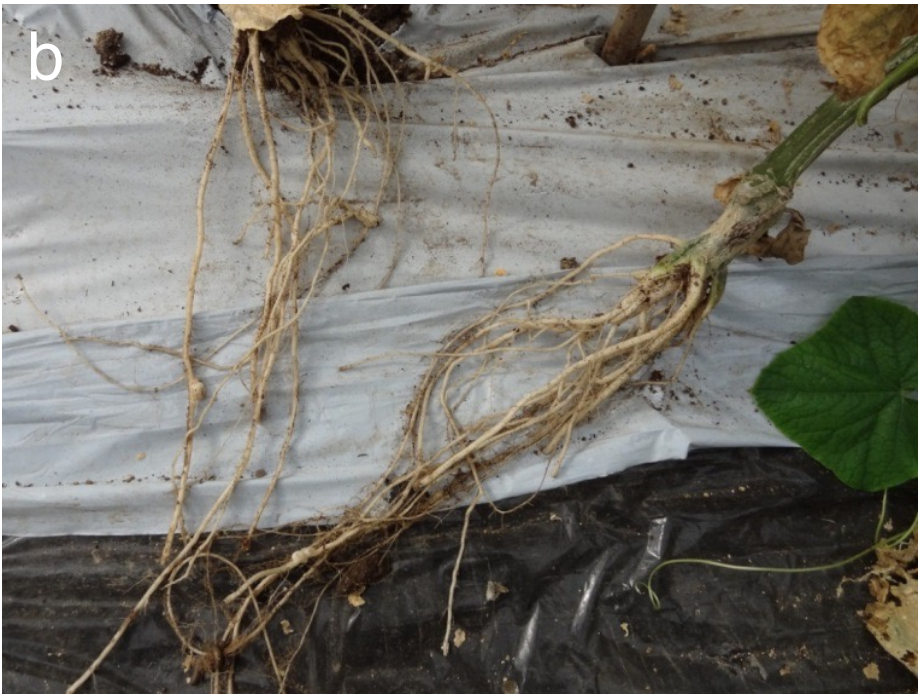

**Fig. S15**
Photos of cucumber roots in a field where damage to the plant parasitic nematode has been observed: (a) roots with several humps infected with the nematode in 2010 (b) roots infected with the nematode but with a few humps in 2011 (c) roots with fewer humps even though the nematodes were present in the soil in 2014. In the field, a 100-fold diluent of thermophile-fermented compost was added to the soil after 2010. The diluent was irrigated in the soil at least once every two weeks during the cultivation period.

## Table S1
Statistical values of the final optimal structural equation models for the metabolites of the leaf and root.

| Category | Model | Fit indices | | |
|---|---|---|---|---|
| No.1 | L_Flavonoid ~ L_Aminoacid | chisq 2.061 | df 4.000 | pvalue 0.725 |
| | L_DPPH ~ L_Flavonoid | cfi 1.000 | tli 1.128 | rfi 0.893 |
| | L_Flavonoid + R_Carotenoid + L_DPPH ~ Compost | nfi 0.952 | srmr 0.033 | AIC 145.889 |
| | lavaan 0.6-11 ended normally after 2 iterations | rmsea 0.000 | gfi 0.997 | agfi 0.988 |
| | Number of successful bootstrap draws 909 | | | |
| No.2 | L_Flavonoid ~ L_Aminoacid | chisq 2.115 | df 5.000 | pvalue 0.833 |
| | L_Flavonoid + R_Carotenoid + L_DPPH ~ Compost | cfi 1.000 | tli 1.152 | rfi 0.912 |
| | | nfi 0.951 | srmr 0.038 | AIC 143.943 |
| | lavaan 0.6-11 ended normally after 1 iterations | rmsea 0.000 | gfi 0.996 | agfi 0.989 |
| | Number of successful bootstrap draws 939 | | | |
| No. 3 | Compost + L_Flavonoid ~ L_Aminoacid | chisq 11.835 | df 8.000 | pvalue 0.159 |
| | L_Other ~ L_Flavonoid | cfi 0.936 | tli 0.879 | rfi 0.702 |
| | L_Flavonoid + R_Carotenoid + L_Other + L_DPPH ~ Compost | nfi 0.841 | srmr 0.049 | AIC 196.609 |
| | lavaan 0.6-11 ended normally after 2 iterations | rmsea 0.219 | gfi 0.853 | agfi 0.613 |
| | Number of successful bootstrap draws 719 | | | |

Th abbreviations in the table indicate the following: L_, leaf; R_, root; chisq, Chi-square: $\chi^2$; df, degrees of freedom (DF); p-value, p values (Chi-square); cfi, comparative fix index (CFI); tli Tucker–Lewis index (TLI); nfi, (non) normed fit index; rfi, relative fit index (RFI); srmr, standardized root mean residuals (SRMR); AIC, Akaike information criterion; rmsea, root mean square error of approximation (RMSEA); gfi, goodness-of-fit index (GFI); and agfi, adjusted goodness-of-fit index (AGFI). Column No. 1 shows the best numerical structural equation model (Fig. 4a). Column No. 2 shows the inferior numerical structural equation model. Column No. 3 shows another inferior numerical structural equation model.

## Table S2
Statistical values of the optimal structural equation model candidates for the characteristic metabolites of the leaf and root.

| Category | Model | Fit indices | | |
|---|---|---|---|---|
| No1 | L_Apigenin_7_O_neohesperidoside + R_Phenylalanine ~ R_2_Aminoadipate + Compost<br>R_2_Aminoadipate ~ Compost<br>L_Apigenin_7_O_neohesperidoside ~ L_Kaempferol + L_Quercetin_Glc<br><br>lavaan 0.6-11 ended normally after 1 iterations<br>Number of successful bootstrap draws        640 | chisq 3.048<br>cfi 1.000<br>nfi 0.943<br>rmsea 0.000 | df 5.000<br>tli 1.112<br>srmr 0.054<br>gfi 1.000 | pvalue 0.693<br>rfi 0.864<br>AIC 45.064<br>agfi 0.999 |
| No2 | L_Apigenin_7_O_neohesperidoside + R_Phenylalanine ~ R_2_Aminoadipate + Compost<br>R_2_Aminoadipate ~ Compost<br>L_Apigenin_7_O_neohesperidoside ~ L_Kaempferol + L_Quercetin_Glc<br>Compost ~ L_Apigenin_7_O_neohesperidoside<br><br>lavaan 0.6-11 ended normally after 33 iterations<br>Number of successful bootstrap draws        637 | chisq 2.857<br>cfi 1.000<br>nfi 0.957<br>rmsea 0.000 | df 6.000<br>tli 1.141<br>srmr 0.046<br>gfi 1.000 | pvalue 0.827<br>rfi 0.899<br>AIC 51.102<br>agfi 0.999 |
| No3 | L_Apigenin_7_O_neohesperidoside + R_Phenylalanine ~ R_2_Aminoadipate + Compost<br>R_2_Aminoadipate ~ Compost<br>L_Apigenin_7_O_neohesperidoside ~ L_Kaempferol<br>Compost ~ L_Apigenin_7_O_neohesperidoside<br><br>lavaan 0.6-11 ended normally after 32 iterations<br>Number of successful bootstrap draws        916 | chisq 2.085<br>cfi 1.000<br>nfi 0.968<br>rmsea 0.000 | df 3.000<br>tli 1.055<br>srmr 0.041<br>gfi 1.000 | pvalue 0.555<br>rfi 0.893<br>AIC 49.316<br>agfi 0.999 |
| No4 | L_Apigenin_7_O_neohesperidoside + R_Phenylalanine ~ R_2_Aminoadipate + Compost<br>R_2_Aminoadipate ~ Compost<br><br>lavaan 0.6-11 ended normally after 1 iterations<br>Number of successful bootstrap draws        992 | chisq 0.241<br>cfi 1.000<br>nfi 0.991<br>rmsea 0.000 | df 1.000<br>tli 1.229<br>srmr 0.018<br>gfi 0.996 | pvalue 0.624<br>rfi 0.944<br>AIC 66.167<br>agfi 0.960 |

The abbreviations in the table indicate the following: L_, leaf; R_, root; chisq, Chi-square: $\chi^2$; df, degrees of freedom (DF); p-value, p values (Chi-square); cfi, comparative fix index (CFI); tli Tucker–Lewis index (TLI); nfi, (non) normed fit index; rfi, relative fit index (RFI); srmr, standardized root mean residuals (SRMR); AIC, Akaike information criterion; rmsea, root mean square error of approximation (RMSEA); gfi, goodness-of-fit index (GFI); and agfi, adjusted goodness-of-fit index (AGFI). Column No. 1 shows the best numerical structural equation model (Fig. 4a). Column No. 2 shows the inferior numerical structural equation model. Column No. 3 shows another inferior numerical structural equation model.

## Table S3
Statistical values of the final optimal structural equation model candidates for the metabolites and bacteria of soil.

| Category | Model | Fit indices | | |
|---|---|---|---|---|
| No1 | Compost ~ S_Methyl_L_cysteine + S_Nicotinamide + Paenibacillus<br>Paenibacillus ~ Compost<br>Compost + S_Nicotinamide ~ S_Aminoadipate<br>S_Nicotinamide ~ S_Methyl_L_cysteine<br>lavaan 0.6-11 ended normally after 40 iterations<br>Number of successful bootstrap draws    24 | chisq 0.989<br>cfi 1.000<br>nfi 0.979<br>rmsea 0.000 | df 2.000<br>tli 1.121<br>srmr 0.005<br>gfi 1.000 | pvalue 0.61<br>rfi 0.905<br>AIC -58.057<br>agfi 1.000 |
| No2 | Compost ~ S_Methyl_L_cysteine + S_Nicotinamide + Paenibacillus<br>Paenibacillus ~ Compost<br>Compost ~ S_Aminoadipate<br>lavaan 0.6-11 ended normally after 38 iterations<br>Number of successful bootstrap draws    42 | chisq 0.989<br>cfi 1.000<br>nfi 0.977<br>rmsea 0.000 | df 2.000<br>tli 1.101<br>srmr 0.055<br>gfi 1.000 | pvalue 0.61<br>rfi 0.918<br>AIC -33.11<br>agfi 1.000 |
| No3 | Compost ~ S_Methyl_L_cysteine + S_Nicotinamide + Paenibacillus + Geobacillus<br>Paenibacillus ~ Compost<br><br>lavaan 0.6-11 ended normally after 21 iterations<br>Number of successful bootstrap draws    60 | chisq 1.128<br>cfi 1.000<br>nfi 0.963<br>rmsea 0 | df 2.000<br>tli 1.128<br>srmr 0.049<br>gfi 1.000 | pvalue 0.569<br>rfi 0.872<br>AIC -21.697<br>agfi 0.999 |
| No4 | Compost ~ S_Methyl_L_cysteine + S_Nicotinamide + Paenibacillus<br>Paenibacillus ~ Compost<br><br>lavaan 0.6-11 ended normally after 31 iterations<br>Number of successful bootstrap draws    310 | chisq 0.983<br>cfi 1.000<br>nfi 0.938<br>rmsea 0 | df 1.000<br>tli 1.008<br>srmr 0.055<br>gfi 0.980 | pvalue 0.322<br>rfi 0.688<br>AIC -8.746<br>agfi 0.799 |

The abbreviations in the table indicate the following: S_, soil; Methyl_L_cysteine, *S*-methyl L-cysteine; chisq, Chi-square: $\chi^2$; df, degrees of freedom (DF); p-value, p values (Chi-square); cfi, comparative fix index (CFI); tli Tucker–Lewis index (TLI); nfi, (non) normed fit index; rfi, relative fit index (RFI); srmr, standardized root mean residuals (SRMR); AIC, Akaike information criterion; rmsea, root mean square error of approximation (RMSEA); gfi, goodness-of-fit index (GFI); and agfi, adjusted goodness-of-fit index (AGFI). Column No. 1 shows the best numerical structural equation model (Fig. 4a). Column No. 2 shows the inferior numerical structural equation model. Column No. 3 shows another inferior numerical structural equation model.

## Table S4
A list of models targeted by causal mediation analysis for Fig. 4a and their statistical values.

| Regression models | (I) L_Flavonoid ~ L_Aminoacid |
| | (II) L_DPPH ~ L_Flavonoid |
| | (III) L_Flavonoid + R_Carotenoid + L_DPPH ~ Compost |

| (I) | Estimate std. | Error | t value | Pr ( > |t|) | | (III) | Estimate std. | Error | t value | Pr ( > |t|) | |
|---|---|---|---|---|---|---|---|---|---|---|---|
| (Intercept) | -8.274 | 3.747 | -2.208 | 0.05825 | # | (Intercept) | 235.87 | 12.8 | 18.425 | 7.75E-08 | *** |
| L_Aminoacid | 22.876 | 6.428 | 3.559 | 0.00742 | ** | Compost | -89.86 | 18.1 | -4.963 | 0.0011 | ** |

Residual standard error: 1.617 on 8 degrees of freedom
Multiple R-squared: 0.6128, Adjusted R-squared: 0.5645
F-statistic: 12.66 on 1 and 8 DF, p-value: 0.007416

Residual standard error: 28.63 on 8 degrees of freedom
Multiple R-squared: 0.7549, Adjusted R-squared: 0.7242
F-statistic: 24.63 on 1 and 8 DF p-value: 0.001102

| (II) | Estimate std. | Error | t value | Pr ( > |t|) | |
|---|---|---|---|---|---|
| (Intercept) | 111.043 | 30.242 | 3.672 | 0.00629 | ** |
| L_Flavonoid | 15.056 | 5.543 | 2.716 | 0.02641 | * |

Residual standard error: 0.004346 on 7 degrees of freedom
Multiple R-squared: 0.4798, Adjusted R-squared: 0.4147
F-statistic: 7.377 on 1 and 8 DF, p-value: 0.02641

The non-parametric bootstrap confidence intervals and Quasi-Bayesian confidence intervals were not calculable. The coefficients and the related calculated data in the regression model are shown on the left. The values mediated among the regression models are shown on the right. The number in "Stimulations shows bootstrapping frequency. The abbreviations in the table indicate the following: T, treat; M, mediator; ACME, average causal mediation (indirect) effect; ADE, average direct effect; Total Effect, mediation (indirect) and direct effect; Prop.Mediated, proportion of mediated effect; #, p <0.1.

## Table S5
A list of models targeted by causal mediation analysis for Fig. 4b and their statistical values.

| Regression models | (I) L_Apigenin_7_O_neohesperidoside + R_Phenylalanine ~ R_2_Aminoadipate + Compost <br> (II) R_2_Aminoadipate ~ Compost <br> (III) L_Apigenin_7_O_neohesperidoside ~ L_Kaempferol + L_Quercetin_Glc |
|---|---|

| (I) | Estimate std. | Error | t value | Pr ( > ltl) | | (III) | Estimate std. | Error | t value | Pr ( > ltl) | |
|---|---|---|---|---|---|---|---|---|---|---|---|
| (Intercept) | 6.956 | 3.086 | 2.254 | 0.0588 | # | (Intercept) | -1.7507 | 0.3591 | -4.875 | 1.80E-03 | ** |
| R_2_Aminoadipate | 3.284 | 2.601 | 1.263 | 0.2472 | | L_Kaempferol | 3.2832 | 0.425 | 7.726 | 0.000114 | *** |
| Compost | -3.551 | 1.359 | -2.612 | 0.0348 | * | L_Quercetin_Glc | -0.3031 | 0.8364 | -0.362 | 0.727785 | |

Residual standard error: 1.603 on 7 degrees of freedom  
Multiple R-squared: 0.7668, Adjusted R-squared: 0.7001  
F-statistic: 11.51 on 2 and 7 DF, p-value: 0.006126

Residual standard error: 0.2459 on 7 degrees of freedom  
Multiple R-squared: 0.975, Adjusted R-squared: 0.9679  
F-statistic:136.5 on 2 and 7 DF p-value: 2.469e-06

| (II) | Estimate std. | Error | t value | Pr ( > ltl) | |
|---|---|---|---|---|---|
| (Intercept) | 1.15388 | 0.09748 | 11.838 | 2.38E-06 | *** |
| Compost | -0.34803 | 0.13785 | -2.525 | 0.0355 | * |

Residual standard error: 0.218 on 8 degrees of freedom  
Multiple R-squared: 0.4434, Adjusted R-squared: 0.3739  
F-statistic: 6.374 on 1 and 8 DF, p-value: 0.03555

The non-parametric bootstrap confidence intervals and Quasi-Bayesian confidence intervals were not calculable. The coefficients and the related calculated data in the regression model are shown on the left. The values mediated among the regression models are shown on the right. The number in "Stimulations shows bootstrapping frequency. Th abbreviations in the table indicate the following: T, treat; M, mediator; ACME, average causal mediation (indirect) effect; ADE, average direct effect; Total Effect, mediation (indirect) and direct effect; Prop.Mediated, proportion of mediated effect; *, p<0.05; **, p<0.01; #, p <0.1.

## Table S6
A list of models targeted by causal mediation analysis for Fig. 4c and their statistical values.

| Regression models | | | | | | |
|---|---|---|---|---|---|---|
| (I) Compost ~ S_Methyl_L_cysteine + S_Nicotinamide + Paenibacillus | | | | | | |
| (II) Paenibacillus ~ Compost | | | | | | |
| (III) Compost + S_Nicotinamide ~ S_Aminoadipate | | | | | | |
| (IV) S_Nicotinamide ~ S_Methyl_L_cysteine | | | | | | |

| (I) | Estimate std. | Error | t value | Pr ( > |t|) | | |
|---|---|---|---|---|---|---|
| (Intercept) | 0.2428 | 1.2684 | 0.191 | 0.866 | | |
| S_Methyl_L_cysteine | 24.4635 | 23.1199 | 1.058 | 0.401 | | |
| S_Nicotinamide | -5.7257 | 11.063 | -0.518 | 0.656 | | |
| Paenibacillus | 2.4642 | 4.6093 | 0.535 | 0.646 | | |

Residual standard error: 0.4327 on 2 degrees of freedom
Multiple R-squared: 0.7503, Adjusted R-squared: 0.3758
F-statistic: 2.004 on 3 and 2 DF, p-value: 0.35

Uncalculable on (I), (II), and (II) since the regression models did not have significant values (Nonparametric bootstrap Confidence Intervals and Quasi-Bayesian Confidence Intervals).

| (II) | Estimate std. | Error | t value | Pr ( > |t|) | |
|---|---|---|---|---|---|
| (Intercept) | 0.1 | 0.03143 | 3.182 | 0.0335 | * |
| Compost | 0.11111 | 0.04444 | 2.5 | 0.0668 | # |

Residual standard error: 0.05443 on 4 degrees of freedom;
Multiple R-squared: 0.6098, Adjusted R-squared: 0.5122
F-statistic: 6.25 on 1 and 4 DF, p-value: 0.06677

Nonparametric bootstrap Confidence Intervals
 (treat: S_Nicotinamide; mediator: S_Methyl_L_cysteine)

| (I) and (IV) | Estimate | 95% CI Lower | 95% CI Upper | p-value |
|---|---|---|---|---|
| ACME | 1.31 | -173.45 | 83.6 | 6.20E-01 |
| ADE | 0 | 0 | 0 | 1 |
| Total Effect | 1.31 | -173.45 | 83.6 | 0.62 |
| Prop. Mediated | 1 | 1 | 1 | NA |

Sample Size Used: 6
Simulations: 1000

| (III) | Estimate std. | Error | t value | Pr ( > |t|) |
|---|---|---|---|---|
| (Intercept) | 1.3825 | 0.7001 | 1.975 | 0.12 |
| S_Aminoadipate | -31.7255 | 25.9552 | -1.222 | 0.289 |

Residual standard error: 0.506 on 4 degrees of freedom
Multiple R-squared: 0.2719, Adjusted R-squared: 0.08993
F-statistic: 1.494 on 1 and 4 DF, p-value: 0.2887

Quasi-Bayesian Confidence Intervals
 (treat: S_Methyl_L_cysteine; mediator: S_Nicotinamide)

| (I) and (IV) | Estimate | 95% CI Lower | 95% CI Upper | p-value |
|---|---|---|---|---|
| ACME | 0 | 0 | 0 | 1.00E+00 |
| ADE | -0.194 | -2.909 | 2.55 | 0.92 |
| Total Effect | -0.194 | -2.909 | 2.55 | 0.92 |
| Prop. Mediated | 0 | 0 | 0 | 1 |

Sample Size Used: 6
Simulations: 1000

| (III) | Estimate std. | Error | t value | Pr ( > |t|) |
|---|---|---|---|---|
| (Intercept) | 1.3825 | 0.7001 | 1.975 | 0.12 |
| S_Aminoadipate | -31.7255 | 25.9552 | -1.222 | 0.289 |

Residual standard error: 0.506 on 4 degrees of freedom
Multiple R-squared: 0.2719, Adjusted R-squared: 0.08993
F-statistic: 1.494 on 1 and 4 DF, p-value: 0.2887

The non-parametric bootstrap confidence intervals and Quasi-Bayesian confidence intervals were not calculable. The coefficients and the related calculated data in the regression model are shown on the left. The values mediated among the regression models are shown on the right. The number in "Stimulations shows bootstrapping frequency. The abbreviations in the table indicate the following: T, treat; M, mediator; ACME, average causal mediation (indirect) effect; ADE, average direct effect; Total Effect, mediation (indirect) and direct effect; Prop.Mediated, proportion of mediated effect; *, $p<0.05$; #, $p<0.1$.

# Table S7
Nif-related genes identified based on genomic data of *Paenibacillus macerans* HMSSN-036.

| Contig No. | KEGG no. | Annotation | Information of annotation | Identity (%) |
|---|---|---|---|---|
| 44 | K04488 | nitrogen fixation protein NifU and related proteins | >ref|WP_036624774.1| nitrogen fixation protein NifU [Paenibacillus macerans] | 89 |
| 84 | K01609 | nitrogen-fixing NifU domain-containing protein | >ref|WP_036620067.1| nitrogen fixation protein NifU [Paenibacillus macerans] | 95 |
| 87 | K02585 | nitrogen fixation protein NifB | >ref|WP_036628315.1| nitrogen fixation protein NifB [Paenibacillus macerans] | 67 |
| 87 | K02585 | nitrogen fixation protein NifB | >gb|KFM95886.1| nitrogenase cofactor biosynthesis protein NifB [Paenibacillus macerans] | 81 |
| 87 | K02596 | nitrogen fixation protein NifX | >ref|WP_036623755.1| nitrogen fixation protein NifX [Paenibacillus macerans] | 79 |
| 87 | K02596 | nitrogen fixation protein NifX | >ref|WP_036623755.1| nitrogen fixation protein NifX [Paenibacillus macerans] | 82 |
| 87 | K02588 | nitrogenase iron protein NifH [EC:1.18.6.1] | >ref|WP_036623760.1| nitrogenase reductase [Paenibacillus macerans] | 94 |
| 87 | K02588 | nitrogenase iron protein NifH [EC:1.18.6.1] | >ref|WP_036623760.1| nitrogenase reductase [Paenibacillus macerans] | 93 |
| 87 | K02587 | nitrogenase molybdenum-cofactor synthesis protein NifE | >ref|WP_036623757.1| nitrogenase iron-molybdenum cofactor biosynthesis protein NifE [Paenibacillus macerans] | 85 |
| 87 | K02587 | nitrogenase molybdenum-cofactor synthesis protein NifE | >ref|WP_036623757.1| nitrogenase iron-molybdenum cofactor biosynthesis protein NifE [Paenibacillus macerans] | 78 |
| 87 | K02587 | nitrogenase molybdenum-cofactor synthesis protein NifE | >ref|WP_036623757.1| nitrogenase iron-molybdenum cofactor biosynthesis protein NifE [Paenibacillus macerans] | 86 |
| 87 | K02592 | nitrogenase molybdenum-iron protein NifN | >gb|KFM95816.1| nitrogenase molybdenum-iron cofactor biosynthesis protein NifN [Paenibacillus macerans] | 79 |
| 87 | K02592 | nitrogenase molybdenum-iron protein NifN | >gb|KFM95816.1| nitrogenase molybdenum-iron cofactor biosynthesis protein NifN [Paenibacillus macerans] | 84 |

**Table S8**
Nif-related genes identified based on genomic data of *Paenibacillus* sp. HMSSN-139.

| Contig No. | KEGG no. | Annotation | Information of annotation | Identity (%) |
|---|---|---|---|---|
| 6 | - | nitrogen-fixing NifU domain-containing protein | >ref|WP_018754084.1| nitrogen-fixing protein NifU [Paenibacillus sanguinis] | 93 |
| 13 | K04488 | nitrogen fixation protein NifU and related proteins | >ref|WP_036624774.1| nitrogen fixation protein NifU Paenibacillus macerans] | 93 |
| 13 | K04488 | nitrogen fixation protein NifU and related proteins | >ref|WP_028538231.1| nitrogen fixation protein NifU [Paenibacillus sp. J14] | 86 |
| 21 | K02585 | nitrogen fixation protein NifB | >gb|KFM95886.1| nitrogenase cofactor biosynthesis protein NifB [Paenibacillus macerans] | 80 |
| 21 | K02585 | nitrogen fixation protein NifB | >ref|WP_036628315.1| nitrogen fixation protein NifB [Paenibacillus macerans] | 71 |
| 21 | K02596 | nitrogen fixation protein NifX | >ref|WP_036623755.1| nitrogen fixation protein NifX [Paenibacillus macerans] | 77 |
| 21 | K02588 | nitrogenase iron protein NifH [EC:1.18.6.1] | >ref|WP_036623760.1| nitrogenase reductase [Paenibacillus macerans] | 93 |
| 21 | K02587 | nitrogenase molybdenum-cofactor synthesis protein NifE | >ref|WP_036623757.1| nitrogenase iron-molybdenum cofactor biosynthesis protein NifE [Paenibacillus macerans] | 83 |
| 21 | K02587 | nitrogenase molybdenum-cofactor synthesis protein NifE | >ref|WP_036623757.1| nitrogenase iron-molybdenum cofactor biosynthesis protein NifE [Paenibacillus macerans] | 86 |
| 21 | K02592 | nitrogenase molybdenum-iron protein NifN | >ref|WP_036623756.1| nitrogenase iron-molybdenum cofactor biosynthesis protein NifN [Paenibacillus macerans] | 81 |

## Table S9

Other functional gene candidates identified based on genomic data of *Paenibacillus macerans* HMSSN-036. The list shows some selected gene candidates associated with nitrogen cycle, phosphate solubilization, and siderophore reaction, amino acid synthesis, and auxin-related functions.

| Contig No. | KEGG no. | Annotation | Information of annotation | Identity (%) |
|---|---|---|---|---|
| 2 | - | ferredoxin | >ref|WP_036627952.1| ferredoxin [Paenibacillus macerans] | 70 |
| 71 | K05337 | ferredoxin | >ref|WP_036624247.1| ferredoxin [Paenibacillus macerans] | 92 |
| 87 | K02586 | nitrogenase molybdenum-iron protein alpha chain [EC:1.18.6.1] | >ref|WP_036623759.1| nitrogenase molybdenum-iron protein subunit alpha [Paenibacillus macerans] | 91 |
| 87 | K02591 | nitrogenase molybdenum-iron protein beta chain [EC:1.18.6.1] | >ref|WP_036623758.1| nitrogenase molybdenum-iron protein subunit beta [Paenibacillus macerans] | 83 |
| 3 | K14487 | GH3 auxin-responsive promoter-binding protein | >ref|WP_036618245.1| hypothetical protein [Paenibacillus macerans] | 98 |
| 16 | K07088 | auxin efflux carrier | >ref|WP_036625350.1| permease [Paenibacillus macerans] | 95 |
| 103 | - | auxin efflux carrier | >ref|WP_036623550.1| aldo/keto reductase [Paenibacillus macerans] | 97 |
| 113 | - | auxin-induced protein PCNT115 | >ref|WP_010348235.1| aldo/keto reductase [Paenibacillus peoriae] | 95 |
| 2 | K01609 | indole-3-glycerol phosphate synthase [EC:4.1.1.48] | >gb|KFN09453.1| indole-3-glycerol phosphate synthase family protein [Paenibacillus macerans] | 98.8 |
| 1 | K01649 | 2-isopropylmalate synthase [EC:2.3.3.13] | >ref|WP_036624250.1| transferase [Paenibacillus macerans] | 99.8 |
| 61 | K03293 | amino acid transporter, AAT family | >ref|WP_036625213.1| GABA permease (4-amino butyrate transport carrier) [Paenibacillus macerans] | 98.2 |
| 189 | K00823 | 4-aminobutyrate aminotransferase [EC:2.6.1.19] | >ref|WP_036626357.1| aminotransferase class III [Paenibacillus macerans] | 78 |
| 2 | K03711 | Fur family transcriptional regulator, ferric uptake regulator | >ref|WP_036621828.1| Fur family transcriptional regulator [Paenibacillus macerans] | 96 |
| 11 | K03306 | inorganic phosphate transporter, PiT family | >ref|WP_036625809.1| inorganic phosphate transporter [Paenibacillus macerans] | 93 |

## Table S10

Other functional gene candidates identified based on genomic data of *Paenibacillus* sp. HMSSN-139.. The list shows some selected gene candidates asociated with nitrogen cycle, phosphate solubilization, and siderophore reaction, amino acid synthesis-related and auxin-related function.

| Contig No. | KEGG no. | Annotation | Information of annotation | Identity (%) |
|---|---|---|---|---|
| 5 | K05337 | ferredoxin | >ref|WP_009222802.1| MULTISPECIES: ferredoxin [Paenibacillus] | 97 |
| 21 | - | nitrogen fixation protein | >ref|WP_036623754.1| hypothetical protein [Paenibacillus macerans] | 54 |
| 21 | K02586 | nitrogenase molybdenum-iron protein alpha chain [EC:1.18.6.1] | >ref|WP_036623759.1| nitrogenase molybdenum-iron protein subunit alpha [Paenibacillus macerans] | 88 |
| 21 | K02591 | nitrogenase molybdenum-iron protein beta chain [EC:1.18.6.1] | >ref|WP_036623758.1| nitrogenase molybdenum-iron protein subunit beta [Paenibacillus macerans] | 85 |
| 34 | K07088 | auxin efflux carrier | >ref|WP_009226239.1| permease [Paenibacillus sp. oral taxon 786] | 95.8 |
| 15 | K01609 | indole-3-glycerol phosphate synthase [EC:4.1.1.48] | >gb|EES72334.1| indole-3-glycerol phosphate synthase [Paenibacillus sp. oral taxon 786 str. D14] | 90.3 |
| 12 | K03293 | amino acid transporter, AAT family | >ref|WP_028539670.1| GABA permease (4-amino butyrate transport carrier) [Paenibacillus sp. J14] | 90.6 |
| 55 | K00823 | 4-aminobutyrate aminotransferase [EC:2.6.1.19] | >ref|WP_036644459.1| aminotransferase class III [Paenibacillus sp. oral taxon 786] | 77 |
| 5 | K01649 | 2-isopropylmalate synthase [EC:2.3.3.13] | >ref|WP_009222800.1| MULTISPECIES: 2-isopropylmalate synthase/homocitrate synthase [Paenibacillus] | 95.1 |
| 2 | K03711 | Fur family transcriptional regulator, ferric uptake regulator | >ref|WP_036621828.1| Fur family transcriptional regulator [Paenibacillus macerans] | 96 |
| 11 | K03306 | inorganic phosphate transporter, PiT family | >ref|WP_036625809.1| inorganic phosphate transporter [Paenibacillus macerans] | 93 |

**Table S11**

Physicochemical indicators in the soil after cultivation of carrots.

| Cat. | pH (H20) | EC mS/cm | CEC meq/100g | CaO mg/100g | MgO mg/100g | K₂O mg/100g | H₃PO₄ mg/100g | NO₃⁻ mg/100g | NH₄ mg/100g | Humus % | PAC | Zn mg/kg | Cu mg/kg | Fe mg/kg | Mn mg/kg | Total_N % | Total_C % |
|---|---|---|---|---|---|---|---|---|---|---|---|---|---|---|---|---|---|
| Control | 6.32 | 0.27 | 15.7 | 438 | 72.1 | 29.5 | 199 | 0.2 | 1.9 | 2.3 | 1011 | 20.6 | 0.25 | 6.21 | 39.5 | 0.14 | 2.56 |
| Compost | 6.68 | 0.96 | 22.7 | 506 | 100 | 46.4 | 175 | 17.7 | 1.3 | 2.7 | 1528 | 22.1 | 0.21 | 4.33 | 35.6 | 0.15 | 2.93 |

The "PAC" in the table shows the data of Phosphate absorption coefficient.

## Table S12
Statistical values of the structural equation models for the amino acids of the leaf and root.

| Category | Model | Fit indices | | |
|---|---|---|---|---|
| No1 | R_Metformin + L_GABA ~ L_Aminoadipate + Compost | chisq 0.3 | df 1.000 | pvalue 0.584 |
| | L_Aminoadipate ~ Compost | cfi 1.000 | tli 1.145 | rfi 0.949 |
| | | nfi 0.991 | srmr 0.02 | AIC -57.55 |
| | lavaan 0.6-11 ended normally after 1 iterations | rmsea 0.000 | gfi 0.999 | agfi 0.993 |
| | Number of successful bootstrap draws   978 | | | |
| No2 | R_Lysine + L_GABA ~ L_Aminoadipate + Compost | chisq 0.009 | df 1.000 | pvalue 0.924 |
| | L_Aminoadipate ~ Compost | cfi 1.000 | tli 1.257 | rfi 0.998 |
| | | nfi 1.000 | srmr 0.05 | AIC 42.096 |
| | avaan 0.6-11 ended normally after 1 iterations | rmsea 0.000 | gfi 1.000 | agfi 1.000 |
| | Number of successful bootstrap draws   988 | | | |
| No3 | R_Lysine + L_GABA ~ R_L_Aminoadipate + Compost | chisq 0.131 | df 1.000 | pvalue 0.718 |
| | R_L_Aminoadipate ~ Compost | cfi 1.000 | tli 1.344 | rfi 0.963 |
| | | nfi 0.994 | srmr 0.017 | AIC 20.070 |
| | lavaan 0.6-11 ended normally after 2 iterations | rmsea 0 | gfi 0.999 | agfi 0.990 |
| | Number of successful bootstrap draws   977 | | | |
| No4 | R_Arginine + L-GABA ~ L-Aminoadipate + Compost | chisq 0.201 | df 1.000 | pvalue 0.654 |
| | R_L_2_Aminoadipate ~ Compost | cfi 1.000 | tli 1.179 | rfi 0.963 |
| | | nfi 0.994 | srmr 0.018 | AIC 46.055 |
| | lavaan 0.6-11 ended normally after 2 iterations | rmsea 0 | gfi 1.000 | agfi 0.995 |
| | Number of successful bootstrap draws   986 | | | |

Abbreviations in the table indicate the following: L_, leaf; R_, root; chisq, Chi-square: $\chi^2$; df, degrees of freedom (DF); p-value, p values (Chi-square); cfi, comparative fix index (CFI); tli Tucker–Lewis index (TLI); nfi, (non) normed fit index; rfi, relative fit index (RFI); srmr, standardized root mean residuals (SRMR); AIC, Akaike information criterion; rmsea, root mean square error of approximation (RMSEA); gfi, goodness-of-fit index (GFI); and agfi, adjusted goodness-of-fit index (AGFI). Column No. 1 shows the best numerical structural equation model (Fig. 4a). Column No. 2 shows the inferior numerical structural equation model. Column No. 3 shows another inferior numerical structural equation model.

# Supplementary Methods

**Title: An agroecological structure model of compost-soil-plant interactions for sustainable organic farming**

*Running head*: Sustainable Agriculture and Compost

Hirokuni Miyamoto*[1,2,3,4], Katsumi Shigeta[5], Wataru Suda[2], Yasunori Ichihashi[6], Naoto Nihei[7], Makiko Matsuura[1,3], Arisa Tsuboi[4], Naoki Tominaga[5], Masahiko Aono[5], Muneo Sato[8], Shunya Taguchi[9], Teruno Nakaguma[1,3,4], Naoko Tsuji[3], Chitose Ishii[2,3], Teruo Matsushita[3,4], Chie Shindo[2], Toshiaki Ito[10], Tamotsu Kato[2], Atsushi Kurotani[8,11], Hideaki Shima[8], Shigeharu Moriya[12], Satoshi Wada[12], Sankichi Horiuchi[13], Takashi Satoh[14], Kenichi Mori[1,3,4], Takumi Nishiuchi[15], Hisashi Miyamoto[3,16], Hiroaki Kodama[1], Masahira Hattori[2,17,18], Hiroshi Ohno[2], Jun Kikuchi*[8], Masami Yokota Hirai*[8]

*Affiliations:*
1. Graduate School of Horticulture, Chiba University, Matsudo, Chiba 271-8501, Japan
2. RIKEN Center for Integrative Medical Science, Yokohama, Kanagawa 230-0045, Japan
3. Sermas Co., Ltd., Ichikawa, Chiba 272-0033, Japan
4. Japan Eco-science (Nikkan Kagaku) Co. Ltd., Chiba, Chiba 260-0034, Japan
5. Takii Seed Co.Ltd., Konan, Shiga 520-3231, Japan
6. RIKEN BioResource Research Center, Tsukuba, Ibaraki 305-0074, Japan
7. Faculty of Food and Agricultural Sciences, Fukushima University, Fukushima, Fukushima 960-1296, Japan
8. RIKEN Center for Sustainable Resource Science, Yokohama, Kanagawa 230-0045, Japan
9. Center for frontier Medical Engineering, Chiba University, Chiba, Chiba 263-8522, Japan
10. Keiyo Gas Energy Solution Co. Ltd., Ichikawa, Chiba 272-0033, Japan
11. Research Center for Agricultural Information Technology, National Agriculture and Food Research Organization, Tsukuba, Ibaraki 305-0856, Japan
12. RIKEN, Center for Advanced Photonics, Wako, Saitama 351-0198 Japan
13. Division of Gastroenterology and Hepatology, The Jikei University School of Medicine, Kashiwa Hospital, Kashiwa, Chiba 277-8567, Japan
14. Division of Hematology, Kitasato University School of Allied Health Sciences, Sagamihara, Kanagawa 252-0329, Japan
15. Division of Integrated Omics research, Bioscience Core Facility, Research Center for Experimental Modeling of Human Disease, Kanazawa University, Kanazawa, Ishikawa, 920-8640, Japan
16. Miroku Co.Ltd., Kitsuki, Oita 873-0021, Japan
17. School of Advanced Science and Engineering, Waseda University, Tokyo169-8555, Japan
18. School of Agricultural and Life Sciences, The University of Tokyo, Bunkyo, Tokyo 113-8657, Japan

\* Hirokuni Miyamoto Ph.D. *Graduate School of Horticulture, Chiba University ; RIKEN Center for Integrative Medical Science; Sermas Co., Ltd.; Japan Eco-science Co. Ltd.*
**Email:** hirokuni.miyamoto@riken.jp
\* Jun Kikuchi, Ph.D. *RIKEN Center for Sustainable Resource Science*
**Email:** jun.kikuchi@riken.jp
\* Masami Hirai, Ph.D. *RIKEN Center for Sustainable Resource Science*
**Email:** masami.hirai@riken.jp

*Cultivation and harvest survey*
Carrot seeds (Takii Seeds Phytorich Series, Kyo Kurenai, Takii Seeds Co., Ltd.) were sown in two rows located 5 cm apart on the cultivation area (0.6 m x 3 m = 18 m$^2$), and the rows were covered with nonwoven fabric after sowing. A slow-release 7-9-7 fertilizer (Sumika Agrotech Co., Ltd.) (120 g/m²), PK 4-15-30-1 (Sumitomo Chemical Co., Ltd.) (40 g/m²), and bitter lime (Sumika Agrotech Co., Ltd.) (100 g/m²) were applied as the basal fertilizer three days before sowing, and additional nitrate-phosphorus-potassium fertilizer (Sumika Agrotech Co., Ltd.) was applied at 50 g/0.75 m² 45 days after sowing. In the compost zone, thermophile-fermented compost powder (Miroku Co., Ltd., and Keiyo Gas Energy Solution Co., Ltd., Japan) [65] was applied at 15 g/m² simultaneously with the basal fertilizer. The chemical properties were as follows[66]: total C (carbon), 38.6 ± 1.9%; total N (nitrogen), 3.6 ± 0.5%; total P (phosphorus), 2.0 ± 0.5%; total K (potassium), 1.0 ± 0.1%; total Ca (calcium), 1.0 ± 0.1%; total Mg (magnesium), 0.7 ± 0.2%; and $H_2O$ (moisture), 16.0 ± 1.7%. The plants were harvested twice, in November and February, and their stem and leaf weight, root weight, root diameter, and root length were measured.

*RGB color image analysis*
The background color of photographs was changed to black for the image analysis of carrots with Adobe Photoshop software (https://www.adobe.com) and Spyder (https://www.python.ambitious-engineer.com/archives/2105) software in Python 3.8 (https://www.python.org). After first changing the background white (red = 255, green = 255, blue = 255) using Photoshop, the white background color was converted to black (red = 0, green = 0, blue =0) by using Spyder. Next, after applying the OpenCV library, the RGB color matrix values of the targeted portion (Fig. 1d) were calculated and extracted using the functions "imread" and "imwrite" for reading and writing images. The averages of the calculated values were used as the data of each carrot in the RGB color image analysis.

*Taste survey*
The harvested carrot roots were randomly sampled and finely ground in a food processor, and the obtained juice was used for the survey. The samples were subjected to taste evaluation via a nondouble-blinded method according to the following indices: Sweet, sweetness; Rich taste, intensity of taste; Immaturity, green odor peculiar to carrots; Flavor, fragrance (scent) of the roots of the carrots.

*Analysis of carotenoids*
In brief, carotenoid extraction was performed with the exception of some minor modifications according to the official protocol (https://www.mext.go.jp/a_menu/syokuhinseibun/1368931.htm). First, an accurately weighed edible sample (1.00 g) of each carrot was chopped into a few pieces, and the pooled samples were placed in a plastic tube (50 mL). Then, 1.0 mL of a 1% (w/v) sodium chloride (NaCl) solution and 15 mL of 3% (w/v) pyrogallol/EtOH were added to the tube, and the sample was well mixed using a Polytron homogenizer (Kinematica, Luzern,

Switzerland) for approximately 1 min. Next, the sample tube washed the homogenizer blade with 5 mL of 3% (w/v) pyrogallol/EtOH. After adding 2.0 mL of a 60% (w/v) potassium hydroxide (KOH) solution to the tube, the mixture was heated at 70°C for 30 min to achieve saponification. After cooling at room temperature, the mixture was divided into two plastic centrifuge tubes (50 mL). Then, 22.5 mL of a 1% (w/v) NaCl solution and 15 mL of hexane/ethyl acetate (9/1) were added to each tube, and carotenoids were extracted by shaking for 10 min. After centrifugation at 3500 rpm for 10 min, the upper layer was collected. Subsequently, 15 mL of hexane/ethyl acetate (9/1) was added to the tube containing the lower layer, and the extraction procedure (i.e., shaking and centrifugation) was performed again. The combined upper layer containing carotenoids was poured into a round-bottom flask, and the solvent was removed by evaporation using a rotary evaporator. Next, hexane/ethyl acetate (9/1) was added to dissolve the residue, and the mixture was transferred to a 10 ml measuring flask. The sample volume was then appropriately adjusted by adding hexane/ethyl acetate (9/1). The carrot extract solutions were stored at -80°C until HPLC analysis.

Lycopene, alpha-carotene, beta-carotene, and canthaxanthin standards for HPLC analysis originally produced by CaroteNature GmbH (Bern, Switzerland) were purchased from Wako Pure Chemical Ind. Ltd. (Tokyo, Japan). High-performance liquid chromatography (HPLC)-grade ethanol (EtOH), methanol (MeOH), and chloroform were provided by Wako Pure Chemical Ind. Ltd. (Tokyo, Japan). The other reagents and solvents used for carotenoid extraction and HPLC analysis were of analytical grade.

An aliquot (1.0 mL) of the extracted sample solution was filtered through a GL Chromatographic Disc 4P, 0.45 μm (GL Sciences Inc., Tokyo, Japan), to remove very small dust particles and insoluble matter, and the filtrate was dried using a centrifugal concentrator. The residue was resolved in 1000 μL of chloroform, and an aliquot (500 μL) was dried using a centrifugal concentrator. Thereafter, the residue was redissolved in 100 μL of chloroform, and an aliquot (90 μL) was mixed with 10 μL of canthaxanthin (0.1 mg/mL), employed as the internal standard material for the present HPLC procedure to avoid any change in solvent volume during autosampler injection, and allow it to be corrected.

The HPLC equipment consisted of a HITACHI-HPLC system (Hitachi High-Tech Corporation, Tokyo, Japan) containing a pump, UV–VIS detector, column oven, and autosampler for sample injection. The employed HPLC procedure was similar to protocols described in the application sheets (LT028 and LT073) of GL Sciences Inc. (Tokyo, Japan). Analyses were performed in an Inertsil ODS-3 column (3 μm, 2.1 mm × 100 mm, GL Sciences Inc., Tokyo, Japan) with the mobile phase described below applied at a 0.2 mL/min flow rate. The mobile phase was a mixture of MeOH and EtOH (45/55). The column oven temperature, UV–VIS detector wavelength, and injection volume were 50°C, 455 nm, and 10 μL, respectively. Under these HPLC analysis conditions, the carotenoid peaks of the carrot samples were well separated. The amount of each carotenoid (i.e., lycopene, alpha-carotene, or beta-carotene) was calculated from the difference in the peak area ratios of an extract sample and a standard sample. The data were corrected based on the recovery of canthaxanthin (the internal standard).

*DPPH radical-scavenging assay*

As previously described [67], a carrot sample was homogenized with ethanol under ice-cold conditions, and the mixture was centrifuged at 2000 × g for 10 min. The upper ethanol layer was evaporated, and the volume of the extract solution was adjusted. Thereafter, 1.0 mL of distilled water, 1.0 mL of 50 mM Tris buffer solution (pH 7.4), and 1.0 mL of 0.1 mM DPPH solution in ethanol were added to the sample solution in the assay tube. The mixture was incubated at 37°C for 20 min, and the absorbance at 517 nm was measured with an iMark microplate reader (Bio–Rad Co., Ltd., USA). Antioxidant activity was calculated from a calibration curve prepared using a set of standard colors obtained by mixing alpha-tocopherol in ethanol solution. Each value was expressed as the alpha-tocopherol equivalent per gram of sample (nmol/g carrot).

*Metabolome analysis*
As previously described [68], the plants were lyophilized and crushed with beads, and a sample of precisely 4 mg (lyophilized weight) was weighed. Subsequently, 1.0 mL of extraction solvent (0.1% formic acid in 80% methanol, including lidocaine (8.4 nmol/L) and 10-camphor sulfonic acid (210 nmol/L) as the internal standard) was added to the sample, and the mixture was centrifuged (1,000 rpm (9,100 g), 1 min). Then, 25 µL of the supernatant of the extract was transferred to a 96-well plate; 225 µL of extraction solvent was added; the samples were shaken, stirred (1,100 rpm, 6 minutes), and centrifuged (2,000 rpm, flashing); and 25 µL of the supernatant was transferred to a 96-well plate. Next, 250 µL of ultrapure water was added to the dry solid (10 minutes), shaken and redissolved (1,100 rpm, 6 minutes), and centrifuged (2,000 rpm, flashing). Then, 120 µL of the supernatant was transferred to a 384-well plate with a filter and centrifuged (2,000 rpm, 5 minutes). Next, plant (leaves and root) analysis was performed using a liquid chromatography (LC)-tandem quadrupole mass spectrometry (MS) system (LC: Acquity UPLC, MS: Xevo TQ-S, Waters). The sample introduction volume was 1 µL. The obtained raw data were selected according to certain conditions (peak area of plant sample > 3,000, peak area of extraction solvent > 1.000), and the data corrected based on the internal standards were used for analysis.

Metabolome analysis of soils was performed as previously described [69]. In brief, soil mixed with 10 ml of sterile water was filtered through a 5A filter (Advantec Co., Ltd.), and the filtrate was lyophilized and crushed with beads. An appropriate amount of extraction solvent was added according to the lyophilized weight. A sample of each mixture of 2.0 mg dry weight was prepared in 1 mL volume. Finally, 1.0 mL of extraction solvent (0.1% formic acid in 80% methanol, including lidocaine (8.4 nmol/L) and 10-camphor sulfonic acid (210 nmol/L) as the internal standard) was added. The mixture was incubated for 2 min, centrifuged (1,000 rpm (9,100 g), 1 min), and stored frozen at -30°C. The supernatant (25 µL) of the extract was transferred to a 96-well plate; 100 µL of the extraction solvent was added; the mixture was shaken, stirred (1,100 rpm, 6 minutes), and centrifuged (2,000 rpm, flashing); and the supernatant (25 µL) was transferred to a 96-well plate. Next, 500 µL of ultrapure water was added to the dry solid (15 min), and the mixture was shaken, redissolved (1,100 rpm, 6 min), and centrifuged (2,000 rpm, flashing). Then, 120 µL of the supernatant was transferred to a 384-well plate with a filter and centrifuged (2,000 rpm, 2 minutes). Soil analysis was performed using an LC-MS system (LC: Nexera X2, MS: LCMS-8050, Shimadzu). The sample introduction volume was 1.5 µL. Peak checks were performed for all compounds. The raw data were corrected based on the internal standard, and the updated data were used for the analysis. These data were visualized by heatmapping and subjected to association analysis and SEM analysis, as described later.

## DNA extraction from soil

The soil was collected after harvesting, and 5 g of each soil sample was added to a 15 mL tube. Five milliliters of sterile water were added, the soil was well suspended by vortexing, and the filtrate was filtered through filter paper (Advantech 5C) (Advantech Co., Ltd., USA) and collected in a new 15 mL tube. DNA was extracted with a QIAGEN Power Soil DNA Mini Kit (QIAGEN Co., Ltd., USA).

## Meta sequence analysis of bacterial 16S rRNA gene sequences

As previously reported [70], the V1-2 region of the bacterial 16S rRNA gene (27fmod-338r) was sequenced according to a previous report[70]. The amplified fragments were sequenced on an Illumina MiSeq system following the manufacturer's instructions. The paired-end reads were merged using the fastq-join program based on overlapping sequences. Reads showing an average quality value of <25 and inexact matches to both universal primers were filtered out. Filter-passed reads were used for further analysis after trimming both primer sequences. The quality filter-passed reads of each sample were rearranged in descending order according to the corresponding quality values and then clustered into OTUs with a 97% pairwise identity cutoff using the UCLUST program, version 5.2.32 (https://www.drive5.com). The taxonomic assignment of each OTU was performed based on similarity searches against the Ribosomal Database Project (RDP) and the National Center for Biotechnology Information (NCBI) genome database using the GLSEARCH program. α-diversity indices of community richness (Chao1) and diversity (Shannon and Simpson) were calculated, and β-diversity indices were estimated via UniFrac analysis with weighted and unweighted principal coordinate analysis (PCoA). All 16S rRNA gene datasets were deposited in the GenBank Sequence Read Archive database as described in Data availability. Phylogenetic trees were constructed on the basis of the mash distance using neighbor-joining method as previously described[71,72].

## Association analyses

Association analysis is a technically established, elementary method of unsupervised learning that is also used for market research as a type of market basket analysis and is applied to achieve an understanding beyond the logic of numbers using relative numbers [73-76]. It is easy to apply when missing values are a characteristic of association analysis. Therefore, it may be a suitable approach when conditions are set such that the identity for classification is different for each categorized layer, for example, when metabolomic data and microbial population are analyzed. It can identify and classify associated components by subjecting them to conditions in which it is challenging to make horizontal comparisons. To predict the components associated with compost, an association analysis was performed as previously reported[73-76]. In brief, association analysis is an elementary way to infer an effect ("target") from a cause ("source"). In this case, the "source" and "target" are represented as x and y, respectively, and the calculation factors for probability (P) are defined as follows:

support $(x \Rightarrow y) = P(x \cap y)$

"support" is P(xy), joint probability (P) of co-occurrence confidence (x ⇒ y) = P(x ∩ y)/P(x)

"confidence" is P(xy)/P(x), conditional probability (P) of occurrence of y after x has occurred.

lift (x ⇒ y) = P(x ∩ y)/P(x) P(y)

"lift" is P(xy)/P(x)P(y), measure of association/independence

A value of > 1 represents a positive association (if the value indicates independence, a value of < 1 represents a negative association).

Here, association rules were determined by using criterion values of support, confidence, and lift ("support = 0.2, confidence = 0.6, maxlen = 2" and "lift > 1.5"). The data combined with all information such as growth, color, taste, metabolites, and bacterial phyla and genera obtained with or without compost addition were used in the analysis. In this study, crop growth, color analysis, questionnaire survey, metabolome, and microbial analyses are performed for each categorized layer, which is suitable for association analysis as described above. To avoid the differences dependent upon the layers, omics data for the analysis were calculated based on the median value (M) within the data and sorted as 0 (< M) and 1 (> M). The growth, color, and taste data are evaluated at a different time from the samples targeted for omics analysis. Additionally, the frequency with the number of samples is different. Therefore, when there was a significant difference in each analysis, the data were set as 0 (low level) and 1 (high level). Therefore, potentially associated components were explored via this analysis. The packages "arules" and "aruleViz" in R software (https://cran.r-project.org) were applied. The systemic network was rendered by Force Atlas with Noverlap in Gephi 0.9.2.

*Covariance Structure Analysis*

Covariance structure analysis/structural equation modeling (SEM) for confirmatory factor analysis (CFA) was conducted using the R software package "lavaan" [76-78]. The analysis codes refer to the website (https://lavaan.ugent.be). As a hypothesis for CFA, the groups selected via correlation, principal component, and association analyses were utilized as component candidates for the latent construct of metabolites and microbiota. The measured values of the metabolites and the relative abundance of the bacterial population were used for the statistical procedure. The treatments with and without compost amendment were set as 0 and 1, respectively. The model hypotheses were statistically estimated via maximum likelihood (ML) parameter estimation with bootstrapping (n = 1000) using the functions 'lavaan' and 'sem'. The "Number of successful bootstrap draws" in the table indicates the number of successful draws after calculation for standard error (1000 estimates as requested bootstrap draws). The model fit was assessed according to the chi-squared p value (p>0.05, nonsignificant), comparative fit index (CFI) (>0.9), Tucker–Lewis index (TLI) (>0.9), goodness-of-fit index (GFI) (>0.95), adjusted goodness-of-fit index (AGFI) (>0.95), root mean square error of approximation (RSMEA) (<0.05), and standardized root mean residual (SRMR) (< 0.08) as indices of good model fit [79]. The path diagrams of the good model were visualized using the package "semPlot" of R software [80]. The analyses of the estimate values were performed concerning (https://www.pu-hiroshima.ac.jp/p/ttetsuji/R/[83]lavaan2sem.html), and the paths were visualized using the R software packages "DiagrammeR", "htmlwidgets", and "webshot" of R software.

*Isolation of nifH-positive thermostable bacteria*

For the isolation of nitrogen-fixing bacteria from compost, succinate-BTB medium (pH 6.8) was prepared as follows: succinic acid (11.6 g), $Na_2MoO_4 \cdot 2H_2O$ (40 mg/mL), $MgSO_4 \cdot 7H_2O$ (0.2 g), NaCl (0.1 g), $CaCl_2 \cdot 2H_2O$ (0.02 g), $Na_2MoO_4 \cdot 2H_2O$ (40 mg/m L) (50μL), $MnSO_4 \cdot H_2O$ (20 mg/mL) (50 μL), biotin (0.2 L), Fe-EDTA MS (1 U) (9.6 mL) bromothymol blue (BTB) (0. 1g/20 mL of EtOH) (1.5mL), yeast extract (20mg), and agar (15 g). The mixture was sterilized after pH adjustment. Direct PCR with a nifH primer set was performed as previously described[81]. In 192 colonies grown in 140 Petri dishes, two strains, *Paenibacillus macerans* HMSSN-036 and *Paenibacillus* sp. HMSSN-139 was selected. The DNA samples from these isolates were extracted by Power Microbial DNA Isolation Kit (MO Bio Laboratories Co., Ltd., USA) and purified by a QIAquik PCR Purification Kit (QIAGEN Co., Ltd., USA). Genome sequencing was performed as previously described[82] and deposited in the GenBank Sequence Read Archive database as described in Data availability. The isolated strains were observed using scanning electron microscopy with some modifications based on an experimental protocol for the spore forming bacteria as previously reported [83,84]. In brief, the preparation of the sample was implemented as follows: the samples of the bacterial solutions cultured on the petri dish were fixed in phosphate-buffered 2% glutaraldehyde and subsequently post-fixed in 2% osmium tetra-oxide for 2 hours in the ice bath. Then, the specimens were dehydrated in graded ethanol and dried by $CO_2$ critical point dry. Finally, dried specimens were coated by osmium plasma ion coater and were submitted to be observed by Scanning electron microscopy (JSM-7500F, JEOL).

### *Stable isotope analysis for nitrogen fixation*

As previously described, a cultivation test with Arabidopsis thaliana ecotypes Landsberg and Columbia was performed as previously described [66]. A commercial horticultural soil medium (Engei Baido; Kureha, Tokyo, Japan) (granules of diameter 2–3 mm) was used as the only soil for cultivation. The isotope test was performed according to the method described by Yano *et al.*[85]. The soil was mixed with compounds included in stable isotopes: 2 kg of soil; 116.6 mg of $K^{15}NO_3$ (Shoko Science Co., Ltd., Japan); 150.15 mg of $(^{15}NH_4)_2SO_4$ (Shoko Science Co., Ltd., Japan). Approximately 25 to 30 seeds were sown in 180 g of the soil medium. The thermophile-fermented compost powder [65] was mixed well with the soil medium in the test group, and the soil was then watered. The pots were watered with distilled water *ad libitum*. The pots were transparently covered, incubated at 4°C for 24 h and vernalized under continuous illumination with a light intensity greater than 3000 lux at 23°C. Most of the seeds germinated after three days. The cover was removed four days after germination, and three seedlings were selected from each pot. These seedlings were watered every two days with approximately 100 mL of distilled water per pot and were grown for 21 days. The grown seedlings and the soils were freeze-dried and used to determine isotope contents. The isotopic composition of nitrogen and the ratio of $^{15}N$ and $^{14}N$ were determined by using a Flash 2000-DELTA$^{plus}$ Advantage ConFlo III System (EA-IRMS; Thermo Fisher Scientific, USA; owned by Shoko Science Co., Ltd., Japan) according to the conventional protocol [76,86-89].

### *Biological assay for isolated Paenibacillus strains*

Biological assays of the production of auxin, the siderophore reaction, and phosphate solubilization were performed as previously reported [90-92]. The auxin assay was performed using a procedure that was partly modified according to a previously described protocol [90] as follows. Sixty milliliters of a mixture of King B broth (tryptone, 1.2 g; $K_2HPO_4$, 0.069 g; $MgSO_4 \cdot 7H_2O$, 0.09 g; glycerol, 0.9 g; L-tryptophan, 0.6 g) and Salkowski color developing solution (50 mL of 35% $HClO_4$; 1 mL of 0.5 M $FeCl_3$) was prepared. After 20 µL of the colony solutions of isolated *Paenibacillus* strains was added to 15 mL of King B broth, the broths were incubated at 80 rpm and 25°C for 48 h. The broths without and with the colony treatment were centrifuged at 10000 rpm and 4°C for 10 min. Two milliliters of supernatant was mixed with 2 mL of the Salkowski color developing solution and 10 µL of phosphate and thereafter shielded from light and allowed to stand for 30 min at room temperature. The absorbance was measured at 530 nm. The CAS assay for siderophore detection was performed with a procedure partly modified according to a previously described protocol [92] as follows. The 10x MM9 solution ($KH_2PO_4$, 3 g; NaCl, 5 g; $NH_4Cl$, 10 g; D.W., 1 L), CAS solution (Mordant Blue 29, 605 mg; DW, 500 mL), cetyltrimethylammonium bromide (HDTMA) solution (HDTMA, 729 mg; D.W., 400 mL), $FeCl_3 \cdot HCl$ solution (1 mM $FeCl_3 \cdot 6H_2O$, 10 mM HCl), and CAS assay agar (10 x MM9 solution, 100 mL; NaOH, 6 g; PIPES, 30.24 g; agar, 15 g; D.W., 750 mL) were prepared. Subsequently, 100 mL of $FeCl_3 \cdot HCl$ solution was added to 500 mL of CAS solution, and 400 mL of HDTMA solution was slowly mixed into the $FeCl_3 \cdot HCl$ and CAS solution. The mixture solution was used as the CAS-HDTMA solution. CAS assay agar was added to CAS-HDTMA solution and sterilized. At approximately 50°C, 10 mL of 20% glucose solution, 1 mL of 1 M $MgCl_2$ solution, and 1 mL of 100 mM $CaCl_2$ solution were mixed in an Erlenmeyer flask, and thereafter, 10 mL of CAS-HDTMA solution was slowly poured along the wall of the flask. The final solution was solidified on the plate for the CAS assay.

The NBRIP medium for the phosphate solubilization assay was prepared according to a previously described protocol[91]. NBRIP cultivation agar medium (glucose, 10 g; $Ca_3(PO_4)_2$, 5 g; $MgSO_4 \cdot 7H_2O$, 0.25 g; $MgCl_2 \cdot 6H_2O$, 2.5 g; KCl, 0.2 g; $(NH_4)_2SO_4$, 0.1 g; agar, 15 g) was prepared. The solution was sterilized and solidified on the plate for the phosphate solubilization assay.

### *Measurement of $N_2O$ from soil*

Fungi derived from banana stalks were cultured on potato dextrose agar (PDA) as previously described [65]. The fungi grown in two Petri dishes were dissolved in 15 mL of sterile water to prepare a fungal solution. Next, 10 mL of the fungal solution was diluted with 190 mL of sterile water. Subsequently, 1% PDA was added to 700 g of soil (Tanemaki Baido) (Takii Seed, Japan) to prepare PDA-containing soil. Finally, 200 mL of the fungal solution was added to the PDA-containing soil. PDA-containing soil without any fungal solution was also included as a negative control. The soils without and with compost were adjusted as follows: 1) 4 mL of distilled water was added, and 2) 4 mL of compost solution was added. A compost solution was prepared by adding 40 mL of sterile water to 4 g of thermophile compost powder, after which the mixture was filtered into a 50-mL Falcon tube with a 100 µm cell strainer (Falcon Co., Ltd., Japan). Pots containing soil were inserted into a transparent sampling bag with two on-off valves (5 L: No. 1-6664-14) (As One Co., Ltd., Japan), and the pots and the bag (the valve side is on the bottom) were placed on a small table at an angle of approximately 30 degrees for seven days. After seven days, a 1-L aluminum bag (AAK-1) (GL Sciences Co., Ltd., Japan) was filled with nitrogen gas.

The bag was then joined to a transparent sampling bag, and the aluminum bag was collected the next day (Condition I). After storage of these pots at 4°C for 1 month, the air inside was released, the bag was closed, and the stopper was closed; 2 h later, a new aluminum bag used for recovery was connected (Condition II). Measurement of the gas concentration in these aluminum bags was performed using a Picarro G5131-i analyzer (Picarro, Santa Clara, California, USA; https://www.sanyo-si.com/products/maker/picarro/, owned and supported by Sanyo Trading Co., Ltd., Japan) according to a previously reported protocol[93-95].

### *Meta-sequence analysis of fungal communities in the soil*

The fungal communities in PDA-containing soil with the fungal solution (soil prepared for the experiment described in "Measurement of $N_2O$ from soil") were determined based on the DNA sequence information as previously described [96]. In brief, DNA from soils without (n=1) and with compost (n=1) selected randomly was extracted, and the fungal ITS1 region was amplified from each replicate with the ITS universal primer sets for fungal organisms established by GENEWIZ. The forward primer contained the sequence "GTGAATCATCGARTC" and the reverse primers contained the sequence "TCCTCCGCTTATTGAT". DNA sequencing was also conducted by GENEWIZ. Inc., Japan. Taxonomic assignments and estimation of relative abundances from sequencing data were performed using the analysis pipeline of the QIIME software package (https://docs.qiime2.org/). All 16S rRNA gene datasets were deposited in the GenBank Sequence Read Archive database as described in Data availability.

### *Causal Mediation Analysis (CMA)*

Individual causal mediation relationships were calculated using the R software package "mediation" [97] based on the tutorial website (https://rpubs.com/Momen/485122). In brief, the R software packages "mediation", "tidyverse", "knitr", and "caret" were used as previously described [76]. Each regression value in the SEMs was calculated by the 'lm' function. In the case of the significant values, the values of the relationships between components as mediators and outcomes were assessed using the 'mediate' function. As previously described, the estimated average causal mediation (indirect) effect (ACME), average direct effect (ADE), and proportion of total effect via mediation (Prop. Mediated) were calculated by quasi-Bayesian confidence intervals and nonparametric bootstrap intervals with 1000 stimulations ('sims=1000') as the numbers of calculations.

### **BayesLiNGAM**

The BayesLiNGAM method [98], which is a Bayesian score-based approach, was applied for the causal structural inference among components in optimal SEMs as previously described. The BayesLiNGAM method was established by the "fastICA" package (https://cran.r-project.org/web/packages/fastICA) of R software. Based on the website information (https://www.cs.helsinki.fi/group/neuroinf/lingam/bayeslingam/), the percentage data calculated by BayesLiNGAM were visualized by the R package "igraph" as previously described [76].